\documentclass[preprintnumbers, floatfix,letterpaper,aps,prd,epsfig,nofootinbib,
twocolumn
]{revtex4-1}
\usepackage{bm,graphicx,dcolumn,epstopdf,epsf, latexsym,mathbbol, amssymb,amsmath,color,slashed, mathrsfs,mathcomp,simplewick, multirow}
\usepackage[center]{subfigure}
\usepackage{multirow}
\usepackage{makecell}
\usepackage[colorlinks,linkcolor=blue,citecolor=blue,urlcolor=blue]{hyperref}

\begin{document}
\allowdisplaybreaks
 \newcommand{\bq}{\begin{equation}}
 \newcommand{\eq}{\end{equation}}
 \newcommand{\bqn}{\begin{eqnarray}}
 \newcommand{\eqn}{\end{eqnarray}}
 \newcommand{\nb}{\nonumber}
 \newcommand{\lb}{\label}
 \newcommand{\f}{\frac}
 \newcommand{\p}{\partial}
\newcommand{\PRL}{Phys. Rev. Lett.}
\newcommand{\PLB}{Phys. Lett. B}
\newcommand{\PRD}{Phys. Rev. D}
\newcommand{\CQG}{Class. Quantum Grav.}
\newcommand{\JCAP}{J. Cosmol. Astropart. Phys.}
\newcommand{\JHEP}{J. High. Energy. Phys.}
%

\title{Polarized primordial gravitational waves in spatial covariant gravities}

\author{Tao Zhu${}^{a, b}$}
\email{zhut05@zjut.edu.cn; Corresponding author}

\author{Wen Zhao${}^{c, d}$}
\email{wzhao7@ustc.edu.cn}

\author{Anzhong Wang${}^{e}$}
 \email{anzhong$\_$wang@baylor.edu}

\affiliation{
${}^{a}$ Institute for theoretical physics and cosmology, Zhejiang University of Technology, Hangzhou, 310032, China \\
${}^{b}$ United Center for Gravitational Wave Physics (UCGWP), Zhejiang University of Technology, Hangzhou, 310032, China \\
${}^{c}$ CAS Key Laboratory for Research in Galaxies and Cosmology, Department of Astronomy, University of Science and Technology of China, Hefei 230026, China \\
${}^{d}$ School of Astronomy and Space Sciences, University of Science and Technology of China, Hefei, 230026, China \\
${}^{e}$ GCAP-CASPER, Physics Department, Baylor University, Waco, 76798-7316, Texas, USA}

\date{\today}

\begin{abstract}

The spatial covariant gravities provide a natural way to including odd-order spatial derivative terms into the gravitational action, which breaks the parity symmetry at gravitational sector. A lot of parity-violating scalar-tensor theories can be mapped to the spatial covariant framework by imposing the unitary gauge. This provides us with a general framework for exploring the parity-violating effects in primordial gravitational waves (PGWs). The main purpose of this paper is to investigate the polarization of PGWs in the spatial covariant gravities and their possible observational effects. To this end, we first construct the approximate analytical solution to the mode function of the PGWs during the slow-roll inflation by using the uniform asymptotic approximation. With the approximate solution, we calculate explicitly the power spectrum and the corresponding circular polarization of the PGWs analytically. It is shown that the new contributions to power spectrum from spatial covariant gravities contain two parts, one from the parity-preserving terms and the other from the parity-violating terms. While the parity-preserving terms can only affect the overall amplitudes of PGWs, the parity-violating terms induce nonzero circular polarization of PGWs, i.e., the left-hand and right-hand polarization modes of GWs have different amplitudes. The observational implications of this nonzero  circular polarization is also briefly discussed.

\end{abstract}


\maketitle

\section{Introduction}
\renewcommand{\theequation}{1.\arabic{equation}} \setcounter{equation}{0}

The inflation which took place at the early Universe has become a dominant paradigm in the standard cosmology \cite{inflation1, inflation2, inflation3, inflation4, inflation5, inflation6}. In this paradigm, primordial density and gravitational-wave fluctuations are created from quantum fluctuations during the inflation process. The former provides primordial seeds for the formation of observed large-scale structure and creates the temperature anisotropy in the cosmic microwave background (CMB), which was already detected by various CMB experiments \cite{Gorski:1996cf, WMAP, Aghanim:2018eyx, Akrami:2018odb}. The primordial gravitational waves (PGWs), on the other hand, also produce distinguishable signatures in both the spectra of the CMB \cite{TTEEBB, TTEEBB1, TTEEBB2, TTEEBB3, TTEEBB4} and the galaxy shaped power spectrum \cite{Schmidt:2012nw, Schmidt:2013gwa, Dodelson:2003bv, Dodelson, Schmidt:2012ne, Chisari:2014xia, Biagetti:2020lpx}. In CMB, the PGWs can produce the TT, EE, BB, and TE spectra, but the TB and EB spectra vanish if the parity symmetry in gravity is respected \cite{TTEEBB, TTEEBB1, TTEEBB2, TTEEBB3, TTEEBB4}. These signatures are important targets of future CMB experiments \cite{SimonsObservatory:2018koc, Hazumi:2019lys, CMB-S4:2016ple, Li:2017drr}.  Similarly,  the PGWs also leave distinct imprints in the B-mode of the galaxy shaped power spectrum but with vanishing E-B correlation due to the parity conservation of the theory \cite{Schmidt:2012nw, Schmidt:2013gwa, Dodelson:2003bv, Dodelson, Schmidt:2012ne, Chisari:2014xia, Biagetti:2020lpx}. It is therefore expected that the future galaxy surveys could also provide invaluable information about the physics of PGWs \cite{Amendola:2016saw, LSST:2008ijt, Biagetti:2020lpx}. 

In most of inflation models that produce PGWs, the theory of general relativity (GR) is assumed to describe the theory of gravity. Due to the parity symmetry of this theory,  the PGWs have two polarization modes which share exactly the same statistical properties and the corresponding inflationary power spectra take the same form. If the parity symmetry is violated, however, the inflationary power spectra of right- and left-handed PGWs can have different amplitudes. The corresponding relative difference between the power spectra of right- and left-handed PGWs measures the level of the parity violation. In CMB, such parity violating effects can induce nonvanishing TB and EB correlation in CMB at large scales and thus the precise measurement of TB and EB spectra could be an important evidence of the parity violation of the gravitational interaction \cite{parity_CMB1, parity_CMB, parity_CMB2, parity_CMB3, parity_CMB4}. It is also proposed that the future ground- and space-based interferometers (such as LIGO/Virgo \cite{Seto:2007tn, Seto:2008sr}, the Big Bang Observer \cite{Seto:2006dz}, LISA and Taiji/Tianqin \cite{Seto:2020zxw, Orlando:2020oko}, etc) are also able to detect or constrain the parity violating effects in the stochastic gravitational-wave background of primordial origin. In addition, parity-violating PGWs also leaves imprints on the large scale structure of the Universe \cite{Masui:2017fzw} and sources nonzero E-B correlation in the galaxy shape power spectrum \cite{Biagetti:2020lpx}. Thus the future galaxy surveys can provide an important approach for testing or constraining the parity violating effects in PGWs \cite{Biagetti:2020lpx, Masui:2017fzw}.  

Theoretically, gravitational parity violation has to somehow modify the theory of GR. This can be achieved by adding some parity-violating terms into the gravitational action of GR. In fact, the gravitational terms with parity violation are ubiquitous in numerous candidates of quantum gravity, such as string theory, loop quantum gravity, and Horava-Lifshitz gravity. One important example is the Chern-Simons modified gravity, which modifies the GR by adding a gravitational Chern-Simons term, arising from string theory and loop quantum gravity \cite{Lue:1998mq, Alexander:2009tp}. This theory has been extended to a chiral scalar-tensor theory by  including the higher derivatives of the coupling scalar field \cite{Crisostomi:2017ugk}. On the other hand, by breaking the time diffeomorphism (or Lorentz symmetry) of the gravitational theory, one can naturally add parity-violating but spatial covariant terms into the gravitational action. This type of parity-violating theories includes Horava-Lifshitz gravities with parity violations \cite{parity_power8, Takahashi:2009wc, Zhu:2013fja, Wang:2017brl} and more generally, the spatial covariant gravities \cite{Gao:2014soa, Gao:2014fra, Gao:2018znj}. Other parity-violating theories, to mention a few,  include Nieh-Yan modified teleparallel gravity \cite{Li:2020xjt, Li:2021wij}, parity-violating symmetric teleparallel gravities \cite{Conroy:2019ibo, Li:2022vtn}, and standard model extension \cite{Kostelecky:2016kfm, Bailey:2006fd, Mewes:2019dhj, Shao:2020shv, Wang:2021ctl}, Holst gravity \cite{Contaldi:2008yz}, etc. 

In all these modified theories, a basic prediction of parity violation is the circular polarization of PGWs, i.e., the left-hand and right-hand polarization modes of GWs propagate with different behaviors. As we also mentioned in the above, such asymmetry between the left- and right-handed modes of PGWs can induce various observational or experimental effects in CMB, stochastic gravitational-wave background, and galaxy-shaped power spectrum. These phenomenological effects have motivated a lot of works in this directions (see Refs. \cite{parity_power8, Takahashi:2009wc, Zhu:2013fja, Biagetti:2020lpx, Qiao:2021fwi, Qiao:2019wsh, Qiao:2019wsh, Orlando:2022rih, Zhang:2022xmm, Chen:2022soq, Cai:2021uup, Fronimos:2021czc, Bartolo:2020gsh, Fu:2020tlw, Hu:2020rub, Mylova:2019jrj, parity_power10, Odintsov:2021kup, Odintsov2, Peng:2022ttg, Kamada:2021kxi, Lyth:2005jf, soda_JHEP, soda_JCAP, soda_PRD, Qiao:2019hkz, Nilsson:2022mzq, Qiao:2022mln} and references therein for example). It is worth noting that the gravitational-wave constraints on the parity violation in gravity have also been extensively explored in the literature by using the gravitational-wave data realized by LIGO/Virgo Collaboration \cite{Gao:2019liu, Zhao:2019xmm, Wang:2020cub, Okounkova:2021xjv, Nishizawa:2018srh, Zhao:2019szi, Wang:2017igw, Wang:2021gqm, Zhao:2022pun, Wang:2020pgu, Niu:2022yhr, Gong:2021jgg, Wu:2021ndf}. 

Spatial covariant gravities is one of modified theory of GR, which breaks the time diffeomorphism of the gravity but respects spatial diffeomorphisms \cite{Gao:2014soa, Gao:2014fra, Gao:2018znj, Gao:2019lpz}. Such spatial covariance provides a natural way to incorporate the parity-violating terms into the theory \cite{Gao:2019liu}. With spatial covariance, the parity violation can be achieved by including the odd-order spatial derivatives into the gravitational action. It is shown in \cite{Gao:2014soa, Gao:2020yzr} that the spatial covariant gravities can provide a unified description for a lot of scalar-tensor theory by imposing the unitary gauge, including those with parity violation, such as the Chern-Simons modified gravity, chiral scalar-tensor theory, Horava-Lifshitz gravities, etc. Therefore, the spatial covariant gravities can provide a general framework for us to explore the parity violating effects in PGWs.  For this purpose, in this paper we study the circularly polarized PGWs in this theory of gravity with parity violation, and the possibility to detect the chirality of PGWs by future potential CMB observations and galaxy surveys.

This paper is organized as follows. In the next section, we present a brief introduction of the construction of the spatial covariant gravities and then discuss the associated propagation of GWs in the a homogeneous and isotropic cosmological background in Sec. III. In Sec. IV, we first derive the master equation that describes the propagation of GWs during inflation and construct the approximate analytical solution to the PGWs by using the uniform asymptotic approximation. With such approximate solution we then calculate explicitly the power spectrum and the polarization of PGWs during the slow-roll inflation. The effects of the parity violation in the CMB spectra and galaxy shaped spectrum, and their detectability have also been briefly discussed. The paper is ended with Sec. V, in which we summarize our main conclusions and provide some outlooks. 

Throughout this paper, the metric convention is chosen as $(-,+,+,+)$, and greek indices $(\mu,\nu,\cdot\cdot\cdot)$ run over $0,1,2,3$ and latin indices $(i, \; j,\;k)$ run over $1, 2, 3$. 

%

\section{spatial covariant gravities}
\renewcommand{\theequation}{2.\arabic{equation}} \setcounter{equation}{0}

In this section, we present a brief introduction of the framework of the spatial covariant gravity, for details about this theory, see \cite{Gao:2014soa, Gao:2014fra} and references therein. 

We first start with the general action of the spatial covariant gravity,
\bqn\lb{action}
S = \int dt d^3 x N \sqrt{g} {\cal L}(N, g_{ij}, K_{ij}, R_{ij}, \nabla_i, \varepsilon_{ijk}),
\eqn
where $N$ is the lapse function, $g_{ij}$ is the 3-dimensional spatial metric, $K_{ij}$ is the extrinsic curvature of $t=$constant hypersurfaces,
\bqn
K_{ij} = \frac{1}{2N} \left(\partial_t g_{ij} - \nabla_i N_j -\nabla_j N_i\right),
\eqn
with $N_i$ being the shift vector, $R_{ij}$ the intrinsic curvature tensor, $\nabla_i$ the spatial covariant derivative with respect to $g_{ij}$, and $\varepsilon_{ijk}=\sqrt{g} \epsilon_{ijk}$ the spatial Levi-Civita tensor with $\epsilon_{ijk}$ being the total antisymmetric tensor. The most important feature of the spatial covariant gravity is that it is only invariant under the three-dimensional spatial diffeomorphism, which breaks the time diffeomorphism. Normally, the violation of the time diffeomorphism can lead to an extra degree of freedom, in addition to the two tensorial degree of freedom in GR. Indeed, it has been verified that the spatial covariant gravity described by the action (\ref{action}) can propagate up to three dynamical degrees of freedom \cite{Gao:2014fra}. In \cite{Gao:2018znj, Gao:2019lpz}, the above action has also been extended by introducing $\dot N$ in the Lagrangian through $\frac{1}{N}(\dot N - N^i \nabla_i N)$. Since such terms does not contribute to the gravitational waves at quadratic order, we will not consider them in this paper.

\begin{table*}
\caption{The building blocks of spatial covariant gravities up to the fourth order in derivatives of $h_{ij}$, where $d_t$, $d_s$ are the number of time and spatial derivative respectively, and $d=d_t+d_s$ denotes the total numbers of time and spatial derivatives. Here $\omega_3(\Gamma)$ denotes the spatial gravitational Chern-Simons term, and $\omega_3(\Gamma) = \varepsilon^{ijk} (\Gamma^m_{jl} \partial_j \Gamma^l_{km} + \frac{2}{3} \Gamma^n_{il} \Gamma^l_{jm} \Gamma^m_{kn})$ with $\Gamma^k_{ij} = \frac{1}{2} g^{km} (\partial_j g_{mj} + \partial_j g_{ij} - \partial_m g_{ij})$ being the spatial Christoffel symbols. The terms in this Table is the same as those of Table. I in \cite{Gao:2019liu} except the two new terms $\omega_3(\Gamma)$ and $\omega_3(\Gamma)K$. }
\lb{blocks}
\begin{ruledtabular}
\begin{tabular} {c|c|c}
$d$  & $(d_t, d_s)$ &  operators\\
 \hline
 0 & $(0, 0)$ & 1 \\
\hline
\multirow{2}*{1} & (1,0)  & $K$ \\
\cline{2-3}
& (0, 1) & - \\
\hline
\multirow{3}*{2} &  (2, 0) & $K_{ij}$, $K^2$ \\
\cline{2-3}
& (1, 1) & - \\
\cline{2-3}
& (0, 2) & $R$ \\
\hline
\multirow{4}*{3} & (3, 0) & $K_{ij} K^{jk} K^i_k$, $K_{ij}K^{ij} K$, $K^3$ \\
\cline{2-3}
& (2, 1) & $\varepsilon_{ijk} K^i_l \nabla^j K^{kl}$ \\
\cline{2-3} 
& (1, 2) & $\nabla^i \nabla^j K_{ij}$, $\nabla^2 K$, $R^{ij} K_{ij}$, $RK$ \\
\cline{2-3}
& (0, 3) & $\omega_3(\Gamma)$ \\
\hline
\multirow{5}*{4} & (4, 0) & $K_{i j} K^{j k} K_{k}^{i} K$, $\left(K_{i j} K^{i j}\right)^{2}$, $K_{i j} K^{i j} K^{2}$,  $K^{4}$ \\
\cline{2-3}
& (3, 1) & $ \varepsilon_{i j k} \nabla_{m} K_{n}^{i} K^{j m} K^{k n}, \quad \varepsilon_{i j k} \nabla^{i} K_{m}^{j} K_{n}^{k} K^{m n}, \quad \varepsilon_{i j k} \nabla^{i} K_{l}^{j} K^{k l} K$ \\
\cline{2-3}
& (2, 2) & $\nabla_{k} K_{i j} \nabla^{k} K^{i j}$, $\nabla_{i} K_{j k} \nabla^{k} K^{i j}$, $\nabla_{i} K^{i j} \nabla_{k} K_{j}^{k}$, $\nabla_{i} K^{i j} \nabla_{j} K$, $\nabla_{i} K \nabla^{i} K$, $R_{i j} K_{k}^{i} K^{j k}$, $R K_{i j} K^{i j}$,  $R_{i j} K^{i j} K $, $RK^{2}$ \\
\cline{2-3}
& (1, 3) & $\varepsilon_{i j k} R^{i l} \nabla^{j} K_{l}^{k}, \quad \varepsilon_{i j k} \nabla^{i} R_{l}^{j} K^{k l}$, $\omega_3(\Gamma) K$ \\
\cline{2-3}
& (0, 4) & $\nabla^{i} \nabla^{j} R_{i j}, \quad \nabla^{2} R, \quad R_{i j} R^{i j}, \quad R^{2}$
\end{tabular}
\end{ruledtabular}
\end{table*}

There are a lot approaches to construct the gravitational theories with spatial covariance. In this paper, we adopt the approach used in \cite{Gao:2019liu} which constructs the Lagrangians of the theory by using the linear combinations of the extrinsic curvature $K_{ij}$, intrinsic curvature $R_{ij}$, as well as their spatial derivatives and derivatives of the spatial metric itself. Then, up to the fourth order in derivatives of $h_{ij}$, we have the building blocks as shown in Table. \ref{blocks} that are all scalars under transformation of spatial diffeomorphisms. Then the general action of the gravitational part will be given by \cite{Gao:2019liu}
\bqn
S_g &=& \int dt d^3 x \sqrt{g} N \Big({\cal L}^{(0)} + {\cal L}^{(1)} + {\cal L}^{(2)} + {\cal L}^{(3)} + {\cal L}^{(4)} \nb \\
&&~~~~~~~~~~~~~~~~~~ ~~~~~~~  + \tilde {\cal L}^{(3)} + \tilde {\cal L}^{(4)}\Big), \lb{oldmodel}
\eqn
where ${\cal L}^{(0)},\; {\cal L}^{(1)}, \;{\cal L}^{(2)}, \; {\cal L}^{(3)}$, and ${\cal L}^{(4)}$ are the parity-preserving terms, which are given by
\bqn
\mathcal{L}^{(0)} &=& c_1^{(0,0)}, \\
\mathcal{L}^{(1)} &=& c_1^{(1,0)} K, \\
\mathcal{L}^{(2)} &=& c_1^{(2, 0)} K_{ij} K^{ij} + c_2^{(2, 0)} K^2 + c_1^{(0, 2)} R, \\
\mathcal{L}^{(3)} &=& c_1^{(3, 0)} K_{ij} K^{jk} K^i_k+ c_2^{(3, 0)} K_{ij}K^{ij} K + c_3^{(3, 0)} K^3 \nb\\
&& + c_1^{(1,2)} \nabla^i \nabla^j K_{ij} + c_2^{(1,2)} \nabla^2 K + c_3^{(1,2)} R^{ij} K_{ij} \nb\\
&& + c_4^{(1,2)} RK, \\
\mathcal{L}^{(4)} &=& c_1^{(4,0)} K_{i j} K^{j k} K_{k}^{i} K + c_2^{(4,0)} \left(K_{i j} K^{i j}\right)^{2} \nb\\
&& + c_3^{(4,0)} K_{i j} K^{i j} K^{2}+ c_4^{(4,0)} K^4 \nb\\
&& + c_1^{(2,2)} \nabla_{k} K_{i j} \nabla^{k} K^{i j} + c_2^{(2,2)} \nabla_{i} K_{j k} \nabla^{k} K^{i j}  \nb\\
&& +c_3^{(2,2)} \nabla_{i} K^{i j} \nabla_{k} K_{j}^{k}+c_4^{(2,2)}\nabla_{i} K^{i j} \nabla_{j} K \nb\\
&&+c_5^{(2,2)}\nabla_{i} K \nabla^{i} K +c_6^{(2,2)} R_{i j} K_{k}^{i} K^{j k}\nb\\
&& +c_7^{(2,2)} R K_{i j} K^{i j} +c_8^{(2,2)} R_{i j} K^{i j} K +c_9^{(2,2)} RK^{2}\nb\\
&&+ c_1^{(0,4)} \nabla^{i} \nabla^{j} R_{i j} + c_2^{(0,4)} \nabla^{2} R  + c_3^{(0,4)} R_{i j} R^{i j} \nb\\
&&+ c_4^{(0,4)}R^2, \lb{4order}
\eqn
and $\tilde{\cal L}^{(3)}$ and $\tilde{\cal L}^{(4)}$ are parity-violating terms which are given by
\bqn
\tilde{\cal L}^{(3)} &=&  c_1^{(2,1)} \varepsilon_{ijk} K^i_l \nabla^j K^{kl} + c_1^{(0,3)} \omega_3(\Gamma),  \lb{L3}\\
\tilde{\cal L}^{(4)} &=& c_1^{(3,1)} \varepsilon_{i j k} \nabla_{m} K_{n}^{i} K^{j m} K^{k n} + c_2^{(3,1)}\varepsilon_{i j k} \nabla^{i} K_{m}^{j} K_{n}^{k} K^{m n} \nb\\
&& + c_3^{(3,1)} \varepsilon_{i j k} \nabla^{i} K_{l}^{j} K^{k l} K + c_1^{(1,3)} \varepsilon_{i j k} R^{i l} \nabla^{j} K_{l}^{k} \nb\\
&& + c_2^{(1,3)} \varepsilon_{i j k} \nabla^{i} R_{l}^{j} K^{k l}+ c_3^{(1,3)} \omega_3(\Gamma) K. \lb{L4}
\eqn
All the coefficients like $c_i^{(d_t, d_s)}$ are functions of $t$ and $N$. Note that in Table.~\ref{blocks} and Eqs.~(\ref{L3}) and  (\ref{L4}), we add the spatial Chern-Simons term $ \omega_3(\Gamma)$ and its coupling to $K$, which are absent in the original action in \cite{Gao:2019liu}. It is interesting to note that the above action reduces to GR if one imposes
\bqn
c_{1}^{(2,0)} = c_1^{(0,2)}=- c_2^{(2, 0)}=\frac{M_{\rm Pl}^2}{2},
\eqn
with all other coefficients $c_{i}^{(d_t, d_s)}$ being setting to zero. 

The spatial covariant gravity described in the above action can represent a very general framework for describing the propagations of GWs in the low-energy effective gravities with Lorentz or parity violation. To our knowledge, a lot of modified gravities can be casted in the framework of the spatial covariant gravity. In addition, it is shown that one in general can relate the spatial covariant gravity to the scalar-tensor theories in the unitary gauge \cite{Gao:2014soa, Gao:2020yzr}.

\section{GWs in spatial covariant gravities}
\renewcommand{\theequation}{3.\arabic{equation}} \setcounter{equation}{0}

Let us investigate the propagation of GWs in the spatial covariant gravities with the action given by (\ref{oldmodel}). We consider the GWs propagating on a homogeneous and isotropic background. The spatial metric in the flat Friedmann- Robertson-Walker universe is written as
\bqn
g_{ij} = a(\tau) (\delta_{ij} + h_{ij}(\tau, x^i)), \lb{metric_spatia}
\eqn
where $\tau$ denotes the conformal time, which relates to the cosmic time $t$ by $dt =a d\tau$, and $a$ is the scale factor of the universe. Throughout this paper, we set the present scale factor $a_0 =1$. $h_{ij}$ denotes the GWs, which represents the transverse and traceless metric perturbations, i.e, 
\bqn
\partial^i h_{ij} =0 = h^i_i.
\eqn
To proceed one can substitute the above spatial metric into the action (\ref{oldmodel}) and expand it to the second order in $h_{ij}$. Here we write the quadratic action in the form as shown in \cite{Gao:2019liu}, 
 \begin{widetext}
 \bqn
 S^{(2)} &=&  \int dtd^3 x \frac{a^3}{2}\Big[ {\cal G}_0(t)\dot h_{ij} \dot h^{ij} + {\cal G}_1(t) \epsilon^{ijk}\dot h_{li} \frac{1}{a} \partial_j \dot h^{l}_k  - {\cal G}_2(t) \dot h_{ij} \frac{\Delta}{a^2} \dot h^{ij}  \nb\\
 &&~~~~~~~~~~ + {\cal W}_0(t) h_{ij} \frac{\Delta}{a^2} h^{ij} + {\cal W}_1(t) \epsilon^{ijk} h_{li} \frac{1}{a} \frac{\Delta}{a^2} \partial_j h^l_k  - {\cal W}_2(t) h_{ij} \frac{\Delta^2}{a^4} h^{ij}\Big], \lb{quadratic_action}\nb\\
 \eqn
 where ${\cal G}_n$ and ${\cal W}_n$ are given by \cite{Gao:2019liu} \footnote{In ${\cal W}_1$ we add the contributions from the two new terms $\omega_3(\Gamma)$ and $\omega_3(\Gamma)K$.}
 \bqn
 {\cal G}_0 &=& \frac{1}{2} \Big[c_1^{(2,0)} + 3 (c_1^{(3,0)} + c_2^{(3, 0)}) H + 3(3 c_1^{(4,0)} + 2 c_2^{(4,0)} + 3 c_3^{(4,0)}) H^2 \Big], \\
 {\cal G}_1 &=& \frac{1}{2} \Big[c_1^{(2,1)}- (c_1^{(3,1)} - 2 c_2^{(3,1)} - 3 c_3^{(3,1)}) H \Big], \\
 {\cal G}_{2} &=& \frac{1}{2}c_1^{(2,2)}, \\
 {\cal W}_0 &=&\frac{1}{2}\Big[ c_1^{(0, 2)} + \frac{1}{2} \dot c_3^{(1,2)} + \frac{1}{2} \Big(3 c_3^{(1,2)} + 6 c_4^{(1,2)} + 2 \dot c_6^{(2,2)} + 3 \dot c_8^{(2,2)}\Big)  H \nb\\
 && + \frac{1}{2}\Big(4c_6^{(2,2)} + 6 c_7^{(2,2)} + 9 c_8^{(2,2)} + 18 c_9^{(2,2)}\Big) H^2 + \frac{1}{2} \Big(2c_6^{(2,2)} + 3 c_8^{(2,2)}\Big)\dot H \Big], \\
 {\cal W}_1 &=& \frac{1}{4} \Big(\dot c_1^{(1,3)} + \dot c_2^{(1,3)}\Big) + c_1^{(0,3)} - 3  c_{3}^{(1,3)} H, \\
 {\cal W}_2 &=& - \frac{1}{2}c_3^{(0,4)}.
 \eqn
  \end{widetext}
 In above a dot denotes the derivative with respect to the cosmic time $t$ and $H=\dot a/a$ is the Hubble parameter. We consider the GWs propagating in the vacuum, and ignore the source term. Varying the action with respect to $h_{ij}$, one can derive the equation of motion for $h_{ij}$ as
 \bqn
 && \left({\cal G}_0 - {\cal G}_2 \frac{\partial^2}{a^2}\right)h''_{ij} + \Big[2 \mathcal{H}{\cal G}_0 + {\cal G}_0' - {\cal G}_2' \frac{\partial^2}{a^2} \Big] h'_{ij} \nb\\
 &&- \left[ {\cal W}_0 - {\cal W}_2 \frac{\partial^2}{a^2} \right]\partial^2 h_{ij} \nb\\
 && + \epsilon_{ilk} \frac{\partial^l}{a} \left[{\cal G}_1 \partial_\tau^2 + ( {\cal H} {\cal G}_1 + {\cal G}_1') \partial_\tau - {\cal W}_1 \partial^2\right] h_{j}^k =0,\nb\\
 \eqn
 where $\mathcal{H} \equiv a'/a$ and a prime denotes the derivative with respect to the conformal time $\tau$.

 \section{Polarization of PGWs}
 \renewcommand{\theequation}{4.\arabic{equation}} \setcounter{equation}{0}
 
 \subsection{Equation of motion for GWs}
 
 In order to study the propagation of GWs in the spatial covariant gravities, it is convenient to decompose the GWs into the circular polarization modes. To study the evolution of $h_{ij}$, we expand it over spatial Fourier harmonics,
 \bqn
 h_{ij}(\tau, x^i) = \sum_{A={\rm R, L}} \int \frac{d^3k}{(2\pi)^3} h_A(\tau, k^i) e^{i k_i x^i} e_{ij}^A(k^i),\nb\\
 \eqn
 where $e_{ij}^A$ denotes the circular polarization tensors and satisfy the relation
 \bqn
 \epsilon^{ijk} n_i e_{kl}^A = i \rho_A e^{j A}_l,
 \eqn
 with $\rho_{\rm R} =1$ and $\rho_{\rm L} = -1$. We find that the propagation equations of these two modes are decoupled, which can be casted into the form \cite{Gao:2019liu}
 \bqn\lb{eom_A}
 h''_A + (2+\Gamma_A) \mathcal{H} h'_A + \omega_A^2 h_A=0,
 \eqn
 where
 \bqn
 {\cal H}\Gamma_A &=& \left[\ln \left({\cal G}_0 +  \rho_A {\cal G}_1 \frac{k}{a} + {\cal G}_2 \frac{k^2}{a^2}\right) \right]',\\
 \frac{\omega_A^2}{k^2} &=& \frac{{\cal W}_0 +  \rho_A {\cal W}_1 \frac{k}{a} + {\cal W}_2 \frac{k^2}{a^2}}{{\cal G}_0 +  \rho_A {\cal G}_1 \frac{k}{a} + {\cal G}_2 \frac{k^2}{a^2}}.
 \eqn
 With this equation, the propagation properties of GWs in the cosmological background have been explored in details in \cite{Gao:2019liu}. Some conditions to make the two polarization modes propagate in the same speed have been considered and a lot of parity-violating gravities with both of polarization modes propagating in the speed of light have been also identified in \cite{Gao:2019liu}. In the above equation, the derivations of the spatial covariant gravities from GR are fully characterized by the quantities $\Gamma_A$ and $\omega_A^2$. The former represents the corrections to the damping rate which modifies the amplitude damping rate of the GWs during their propagations in the cosmological background, and the latter is the modified dispersion relation of GWs which leads to a phase shifting of GWs from distant sources.  
 
 For later convenience of calculating the primordial power spectra of GWs, let us introduce a new variable
 \bqn
 u_A = \sqrt{2}  z h_A,
 \eqn
 with 
 \bqn
 z = a \sqrt{{\cal G}_0 +  \rho_A {\cal G}_1 \frac{k}{a} + {\cal G}_2 \frac{k^2}{a^2}}.
 \eqn
 Then the equation of motion (\ref{eom_A}) can be rewritten in the form
 \bqn
 u''_A + \left( \omega_A^2 - \frac{z''}{z}\right)u_A=0, \lb{eom_uA}
 \eqn
 and we expect the derivations from GR are small such that
\bqn
\Gamma_A \ll 1, \;\; \left|\frac{\omega_A^2}{k^2} -1\right| \ll 1.
\eqn
Thus we can consider all the new effects on GWs beyond GR as small corrections to the standard GR result. In this way, we are able to expand  $\omega_A$ and $\frac{z''}{z}$ as
\bqn
\frac{\omega_A^2}{k^2}& \simeq& \frac{{\cal W}_0}{{\cal G}_0} +  \rho_A \frac{{\cal W}_1 - {\cal G}_1}{{\cal G}_0} \frac{k}{a} + \frac{{\cal W}_2 - {\cal G}_2}{{\cal G}_0} \frac{k^2}{a^2},  \lb{omega}\\
\frac{z''}{z} &\simeq& \left(1-\frac{1}{2} \rho_A \frac{{\cal G}_1}{{\cal G}_0}\frac{k}{a} - \frac{{\cal G}_2}{{\cal G}_0} \frac{k^2}{a^2}\right)\frac{a''}{a} \nb\\
&& + \frac{1}{2} \left(\frac{{\cal G}''_0}{{\cal G}_0} + \rho_A \frac{{\cal G}''_1}{{\cal G}_0} \frac{k}{a} +\frac{{\cal G}''_2}{{\cal G}_0} \frac{k^2}{a^2} \right) \nb\\
&&+ \left(\frac{{\cal G}'_0}{{\cal G}_0} - \frac{{\cal G}'_2}{{\cal G}_0} \frac{k^2}{a^2}\right) \frac{a'}{a} + \frac{{\cal G}_2}{{\cal G}_0} \frac{k^2}{a^2} \frac{a'^2}{a^2}.\lb{zz}
\eqn
Note that in the above expansion, we only consider the first-order terms of each coefficients, i..e, $1-{\cal W}_0/{\cal G}_0$, ${\cal W}_1$, ${\cal G}_1$, ${\cal W}_2$, ${\cal G}_2$, ${\cal G}'_0$, and ${\cal G}''_0$.

In this article, we consider the PGWs during the inflationary stage, and assume that the background evolution during the inflation is slowly varying. With this consideration, we can treat all the coefficients ${\cal G}_0$, ${\cal G}_1$, ${\cal G}_2$, ${\cal W}_0$, ${\cal W}_1$, and ${\cal W}_2$ as slowly varying quantities. Then one is able to expand the modified dispersion relation $\omega_A^2$ in (\ref{omega}) and effective time-dependent mass term $z''/z$ in (\ref{zz}) in terms of the slow-roll quantities as
\bqn
\frac{\omega_A^2}{k^2} &\simeq & \frac{{\cal W}_0}{{\cal G}_0} - \rho_A \frac{{\cal W}_1- {\cal G}_1}{{\cal G}_0} H k \tau \nb\\
&& + \frac{{\cal W}_2 - {\cal G}_2}{{\cal G}_0} H^2 k^2 \tau^2, \lb{omega_sl}
\eqn
and
\bqn\lb{z_sl}
\frac{z''}{z} &\simeq& \frac{1}{\tau^2} \left(2 + 3 \epsilon_1 + \frac{3 H \dot{\cal G}_0+\ddot{\cal G}_0}{2 H^2 {\cal G}_0}\right) \nb\\
&& + \frac{k}{\tau} \rho_A \frac{2H^2 {\cal G}_1 - H \dot{\cal G}_1 - \ddot{\cal G}_1}{2 H {\cal G}_0} \nb\\
&& - k^2 \frac{2 H^2 {\cal G}_2 + H \dot{\cal G}_2 - \ddot{\cal G}_2}{2 {\cal G}_0}.
\eqn
It is worth noting that, in order to obtain the above expansions, we have used the relation
\bqn
a&\simeq &-\frac{1+\epsilon_1}{\tau H},
\eqn
with $\epsilon_1 = - \dot H/H^2$. 

With the expressions of $z''/z$ and $\omega_A^2/k^2$, one observes that the equation of motion in Eq.(\ref{eom_uA}) can be casted into the form
\bqn\lb{zz}
&&u''_A+\Bigg\{-\frac{v_t^2-\frac{1}{4}}{k^2\tau^2} - \rho_A \frac{d_{-1}}{k\tau} + d_0   \nb\\
&&~~~~~~~ - \rho_A d_1 k \tau + d_2 k^2 \tau^2 \Bigg\} k^2 u_A=0,
\eqn
where
\bqn
\nu_t^2& \equiv& \frac{9}{4}+3 \epsilon_1 +  \frac{3 H \dot{\cal G}_0+\ddot{\cal G}_0}{2 H^2 {\cal G}_0} , \\
d_{-1} &\equiv& \frac{2H^2 {\cal G}_1 - H \dot{\cal G}_1 - \ddot{\cal G}_1}{2 H {\cal G}_0}, \\
d_0 &\equiv& \frac{{\cal W}_0}{{\cal G}_0} +  \frac{2 H^2 {\cal G}_2 + H \dot{\cal G}_2 - \ddot{\cal G}_2}{2 {\cal G}_0}, \\
d_1 & \equiv&  \frac{{\cal W}_1- {\cal G}_1}{{\cal G}_0} H ,  \\
d_2 &\equiv&   \frac{{\cal W}_2 - {\cal G}_2}{{\cal G}_0} H^2,
\eqn
and all these coefficients are slowly varying and dimensionless. Obviously,  this equation does not have analytical solutions even if one treats all the slowly varying quantities as constants. In order to obtain its solution, we have to consider some approximations. In this paper, we will consider the uniform asymptotic approximation, which is developed in a series of papers for a better treatment to equations with turning points and poles. This approximation has been widely applied in calculating primordial spectra for various inflation models \cite{zhu_constructing_2014, zhu_inflationary_2014, zhu_power_2014, qiao_inflationary_2018,  habib_inflationary_2002} and applications in studying the reheating process \cite{Zhu:2018smk} and quantum mechanics \cite{Zhu:2019bwj}.  In the following subsections, we apply this approximation to construct the approximate solution of (\ref{zz}) and calculate the corresponding primordial tensor power spectra in the spatial covariant gravities. 

\subsection{Uniform asymptotic approximation}

In this subsection, we employ the uniform asymptotic approximation method to construct the approximate asymptotic solutions. Most of the expressions and results used here can be found in \cite{zhu_constructing_2014, zhu_inflationary_2014, zhu_power_2014, qiao_inflationary_2018, Qiao:2019hkz}.

In the uniform asymptotic approximation, it is convenient to write the equation of motion (\ref{zz}) in the following standard form \cite{olver1954, zhu_constructing_2014, zhu_inflationary_2014},
\bqn
\frac{d^2u_A(y)}{dy^2}=[g(y)+q(y)]u_A(y), \lb{standard}
\eqn
where $y\equiv-k\tau$ is a dimensionless variable and
\bqn
g(y)+q(y)\equiv\frac{v_t^2-\frac{1}{4}}{y^2}+\frac{\rho_A d_{-1}}{y}- d_0 + \rho_A d_1 y - d_2 y^2. \lb{standard1}\nb\\
\eqn
In general, $g(y)$ and $q(y)$ have two poles (singularities): one is at $y=0^+$ and the other is at $y=+\infty$. In the uniform asymptotic approximation, in order to make the approximate solution being valid around the poles, one has to ensure that the error control function associated with the approximate solution to be convergent. For the equation of motion in (\ref{standard}) with $g(y)+q(y)$ given by (\ref{standard1}), it is shown in \cite{zhu_inflationary_2014} that in order to make its error control function to be convergent around the second-order pole at $y=0^+$,  one has to choose,
\bqn\lb{qofy}
q(y)&=&-\frac{1}{4y^2}.
\eqn
Then $g(y)$ is given by
\bqn
g(y)&=&\frac{\nu_t^2}{y^2} + \rho_A \frac{d_{-1}}{y}- d_0 + \rho_A d_{1}y- d_2 y^2. \lb{gofy}
\eqn
Except for the two poles at $y = 0^+$ and $y = +\infty$, $g(y)$ may also have a single zero in the range $y \in (0, +\infty)$, which called a single turning point of $g(y)$. Since in GR limit, we have $\nu_t^2 \simeq \frac{9}{4} + 3 \epsilon_1$, $d_0 = 1$, and $d_{-1} =0= d_1 =d_2$, we expect all the new terms with coefficients $d_{-1}$, $d_1$, and $d_2$ can be considered as small corrections. With this consideration and solving the equation $g(y) = 0$, we obtain the turning point,
\bqn
y_0^A& \simeq & \frac{\nu_t}{\sqrt{d_0}} + \frac{\rho_A d_{-1} + \rho_A  d_1 \nu_t^2 - d_2 \nu_t^3}{2}.
\eqn
In deriving the above expression, we have directly used $d_0=1$ in the second term since $d_0 -1  \simeq \frac{{\cal W}_0}{{\cal G}_0}-1 +  \frac{2 H^2 {\cal G}_2 + H \dot{\cal G}_2 - \ddot{\cal G}_2}{2 {\cal G}_0}$ is a small corrections as well. In the uniform asymptotic approximation, the approximate solution depends on the type of turning point. The turning point $y_0$ is a single root of the equation $g(y)=0$, which also called single turning point as well. Thus, in the following discussion, we will discuss the solution around this single turning point in details.

\begin{figure*}
  \centering
  \includegraphics[width=8cm]{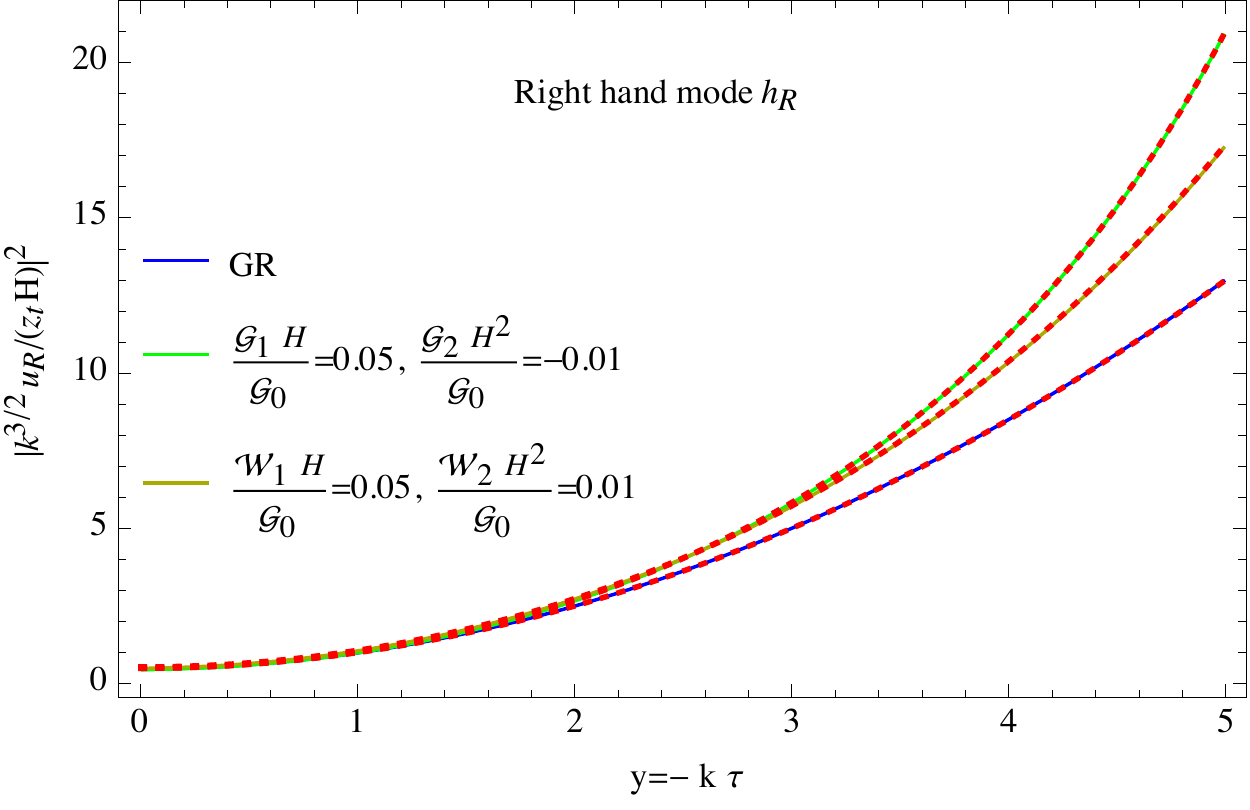}
    \includegraphics[width=8cm]{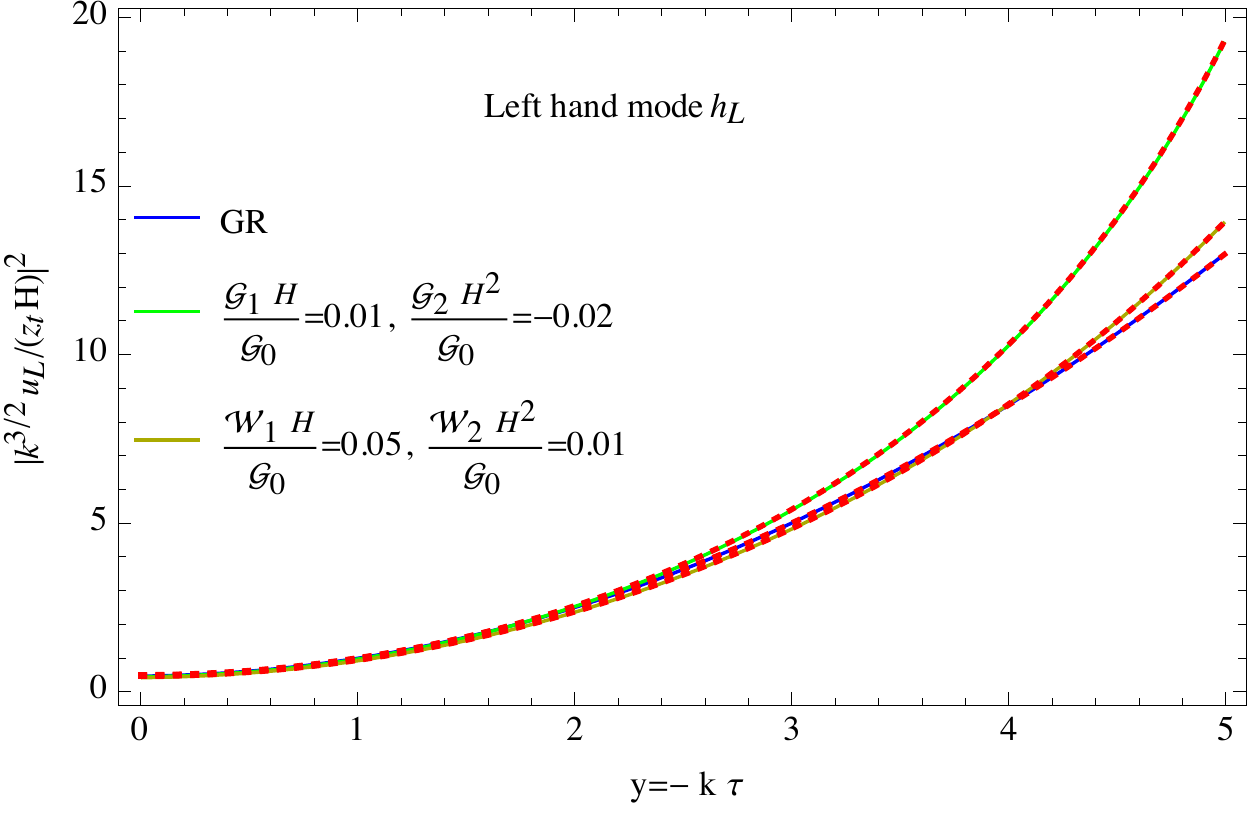}
  \caption{The uniform asymptotic approximate solutions of mode functions $|k^{3/2}u_{A}/(z_t H)|^2$ (solid curves) and the corresponding numerical solutions (dotted curves). Left panel presents the solution of the left hand mode while the right panel presents the right hand mode. In each panel, the solid curves with blue color corresponds to the solution for case of general relativity, and green and darker yellow colors correspond to the spatial covariant gravities with different values of the parity-violating parameters.  The numerical solution associated with each analytical solutions are presented by the red dotted curves.} \label{solution}
\end{figure*}

For the single turning point $y_0$, the approximate solution of equation of motion around this turning point can be expressed in terms of Airy-type functions as \cite{zhu_inflationary_2014}
\bqn\lb{ALY}
u_A=\alpha_0\left(\frac{\xi}{g(y)}\right)^{1/4}{\rm Ai}(\xi)+\beta_0\left(\frac{\xi}{g(y)}\right)^{1/4}{\rm Bi}(\xi),\nb\\
\eqn
where $\rm{Ai}(\xi)$ and $\rm{Bi}(\xi)$ are the Airy functions, $\alpha_0$ and $\beta_0$ are two integration constants, $\xi$ is the function of $y$ and the form of $\xi(y)$ is given by \cite{zhu_inflationary_2014}
\bqn
\xi(y) =
\begin{cases}
\left(-\frac{3}{2}\int^y_{y_0}\sqrt{g(y')}dy'\right)^{2/3} ,\;  & y\leq y_0,\\
-\left(\frac{3}{2}\int^y_{y_0}\sqrt{-g(y')}dy'\right)^{2/3} ,\; & y\geq y_0.\\
\end{cases}\nb\\
\eqn
As shown in \cite{olver1954, zhu_inflationary_2014}, the above solution is not only valid around the turning point, but valid in the while domain $y \in (0, + \infty)$. It is shown in \cite{olver1954, zhu_inflationary_2014} that with the choice of $q(y)$ given in (\ref{qofy}), the error control function of the approximate solution of (\ref{ALY}) is convergent even around the two poles $y=0^+$ and $y = +\infty$.
With this solution, we need to determine the coefficients $\alpha_0$ and $\beta_0$ by matching it with the two boundary conditions. One requires the mode function $u_A$ satisfies the following normalization condition, i.e., 
\bqn
\frac{i}{\hbar} (u_A^* u_A' - u^{*}_A{}' u_A) =1,
\eqn
where $u_A^*$ denotes the complex conjugate of the mode function $u_A$. The second boundary condition that fixes the mode function $u_A$ completely comes from the initial condition in the limit $y \to +\infty$, which corresponds to the  assumption that the universe was initially in an adiabatic vacuum,
\bqn
\lim_{y \to +\infty} u_k(y)&\simeq&\frac{1}{\sqrt{2\omega_A}}e^{-i\int\omega_k d\tau}\nb\\
&=&\sqrt{\frac{1}{2k}}\left(\frac{1}{-g}\right)^{1/4}\exp\left(-i\int^y_{y_i}\sqrt{-g}dy\right).\nb\\
\eqn
When $y\rightarrow+\infty$, we note that $\xi(y)$ is very large and negative. In this limit, the asymptotic form of the Ariy functions read
\bqn
{\rm Ai}(-x)&=&\frac{1}{\pi^{1/2}x^{1/4}}\cos\left(\frac{2}{3}x^{3/2}-\frac{\pi}{4}\right),\\
{\rm Bi}(-x)&=&-\frac{1}{\pi^{1/2}x^{1/4}}\sin\left(\frac{2}{3}x^{3/2}-\frac{\pi}{4}\right).
\eqn
Combining the initial condition and the approximate analytical solution, we obtain
\bqn
\alpha_0=\sqrt{\frac{\pi}{2k}}e^{i\frac{\pi}{4}},~~~
\beta_0=i\sqrt{\frac{\pi}{2k}}e^{i\frac{\pi}{4}}.
\eqn
In Fig.~\ref{solution}, we plotted the time evolution of the power spectra  $|k^{3/2}u_{A}/(z_t H)|^2$ for both the uniform asymptotic solutions and numerical solutions of the right-handed and the left hand modes respectively. We also displayed the cases in GR for comparison. From this figure, one can see clearly that our analytical solutions are extremely close to the numerical ones, and even are not distinguishable from the numerical ones.

\subsection{Power spectra and circular polarization of PGWs}

With the above approximate solutions of the PGWs, we are able to calculate the corresponding primordial power spectra for each polarization modes of the PGWs in the limit $y\to 0$. Their power spectra is normally computed via
\bqn
\mathcal{P}_{\rm T}^{\rm L} = \frac{2 k^3}{\pi^2} \left|\frac{u_{\rm L}(y)}{z}\right|^2, \nb\\
\mathcal{P}_{\rm T}^{\rm R} = \frac{2 k^3}{\pi^2} \left|\frac{u_{\rm R}(y)}{z}\right|^2.
\eqn
In the limit $y\to 0^+$, the variable $\xi(y)$, which is the argument of the Airy function, becomes very large and positive, allowing the use of the following asymptotic forms
\bqn\lb{AB}
{\rm Ai}(x)&=&\frac{1}{2\pi^{1/2}x^{1/4}}\exp\left(-\frac{2}{3}x^{3/2}\right),\\
{\rm Bi}(x)&=&-\frac{1}{\pi^{1/2}x^{1/4}}\exp\left(\frac{2}{3}x^{3/2}\right).
\eqn
These asymptotic forms indicate that only the growing mode of $u_A(y)$ is relevant in the limit $y\to 0$. Thus the approximate solution near the pole $y = 0^+$ can be expressed in the form
\bqn
u_A(y)&\approx&\beta_0\left(\frac{1}{\pi^2g(y)}\right)^{1/4}\exp\left(\int^{y_0}_ydy\sqrt{g(y)}\right)\nb\\
&=&i\frac{1}{\sqrt{2k}}\left(\frac{1}{g(y)}\right)^{1/4}\exp\left(\int^{y_0}_ydy\sqrt{g(y)}\right).\nb\\
\eqn
The power spectra of PGWs are then given by
\bqn\lb{PS}
\mathcal{P}^A_{\rm T}
&=&\frac{k^2}{\pi^2}\frac{1}{z^2}\frac{y}{\nu_t}\exp\left(2\int^{y_0}_ydy\sqrt{g(y)}\right)\nb\\
&\simeq&\frac{H^2}{18 \pi^2 e^{3} {\cal G}_0}\Bigg[1+ \left(2 \ln 2-\frac{8}{3}\right)\epsilon_1 + \frac{(3 \ln 2 -1)\dot{\cal G}_0}{3 H {\cal G}_0}\nb\\
&&~~+ \frac{\pi\rho_A (9d_1 + 8d_{-1})}{16} - \frac{9}{4}d_2\Bigg].
\eqn
Note that in estimation the above integral, we take the limit $y \to 0^+$ and the detail calculation of the integral of $\sqrt{g}$ is presented in Appendix A. Obviously, the power spectra can be modified due to the presence of both the parity-preserving terms and parity-violating terms in the gravitational action (\ref{oldmodel}). It is easy to check that when one takes ${\cal G}_0= M_{\rm Pl}^2/4$ and $d_{-1}=d_1=d_2=0$, the standard GR result can be recovered. The parity-preserving terms can only affect the overall amplitude of both the left- and right-handed polarization modes of GW, which are related to the quantities ${\cal G}_0$ and $d_2$ in the above expression. The relevant terms in the gravitational action are those with coefficients $c_{1}^{(2,0)}$, $c_1^{(3, 0)}$, $c_2^{(3,0)}$, $c_1^{(4,0)}$, $c_2^{(4,0)}$, $c_3^{(4,0)}$, $c_{1}^{(2,2)}$, and $c_3^{(0,4)}$. The parity-violating terms, on the other hand, affect the amplitudes of  left- and right-handed polarization modes of GW in different ways. For positive value of $9d_1+8d_{-1}$ in the above expression, the parity violation trends to enhanced (suppress) the power spectra of the right (left) -handed modes. This effect is related to those terms with coefficients $c_{1}^{(2,1)}$, $c_1^{(3,1)}$, $c_2^{(3,1)}$, $c_3^{(3,1)}$, $c_1^{(1,3)}$, $c_2^{(1,3)}$, $c_2^{(1,3)}$, $c_1^{(0,3)}$, and $c_3^{(1, 3)}$ in the gravitational action (\ref{oldmodel}).

Here we would like to mention that in the calculation of the power spectra, we only consider the first-order approximation in the uniform asymptotic approximation. The corresponding relative error of the overall amplitude $\frac{H^2}{18 \pi^2 e^{3} {\cal G}_0}$ of the power spectra in Eq.~(\ref{PS}) is less than 10\%, see the discussion about the relative error at each order in ref.~\cite{zhu_power_2014}. In principle, this calculation can be significantly improved by considering high-order uniform asymptotic approximation. For example, as shown in \cite{zhu_power_2014}, at the third-order uniform asymptotic approximation, the relative error of the the overall amplitude can be improved to be less than 0.15\%. However, the small corrections presented in the square bracket in (\ref{PS}) can be quit precise provided that these corrections are sufficient small. As we will mentioned later, the resulted circular polarization calculated from  (\ref{PS}) can be exactly reduced to the exact result in the Chern-Simons gravity.

\begin{figure*}
  \centering
  \includegraphics[width=8cm]{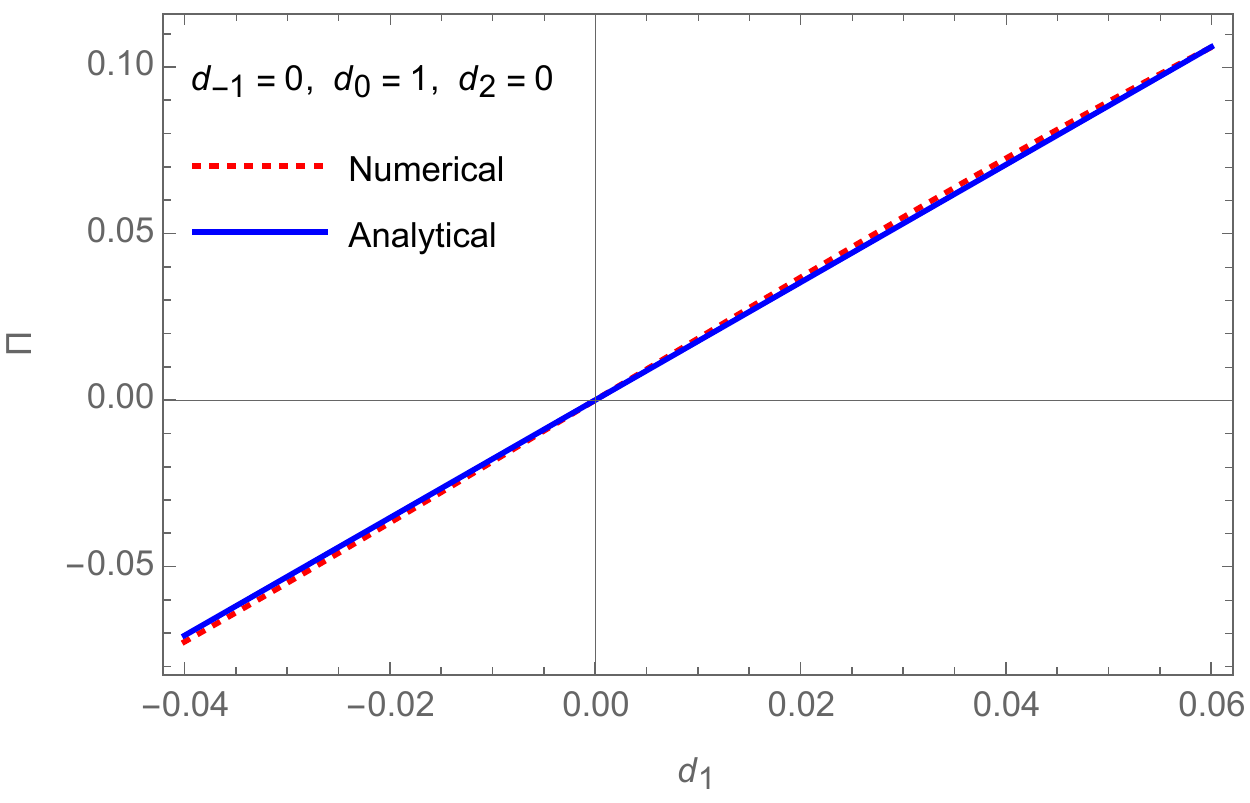}
    \includegraphics[width=8cm]{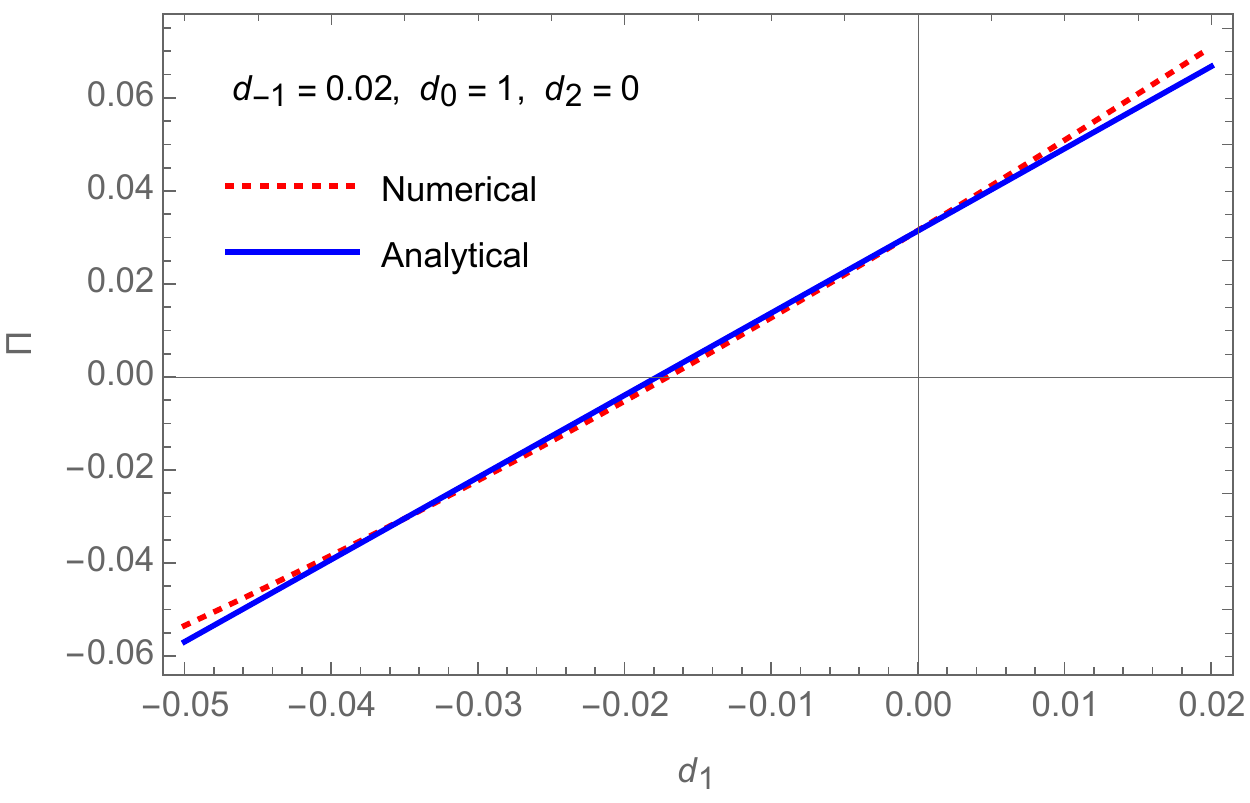}
  \caption{The degree of circular polarization $\Pi$ as a function of $d_1$ for different values of $d_{-1} =0$ (left panel) and $d_{-1}=0.02$ (right panel), respectively. In both panels, the blue and solid curve represent the analytical results in Eq.~(\ref{Pi_ana}) and the red and dotted curves are the numerical results. } \label{Pi_plot}
\end{figure*}

Now, we are in a position to calculate the degree of the circular polarization of PGWs, which is defined by the differences of the amplitudes between the two circular polarization states of PGWs as
\bqn
\Pi&\equiv&\frac{\mathcal{P}^{\rm R}_{\rm T}-\mathcal{P}^{\rm L}_{\rm T}}{\mathcal{P}^{\rm R}_{\rm T}+\mathcal{P}^{\rm L}_{\rm T}} \simeq \frac{\pi (9d_1 + 8d_{-1})}{16}.\lb{Pi_ana}
\eqn
As expected, the degree of the circular polarization $\Pi$ only depends on the parity-violating terms in the gravitational action. It is not difficult to check that the above expression can exactly reduce to the cases of Chern-Simons modified gravity \cite{soda_JCAP}, chiral scalar-tensor theory \cite{Qiao:2019hkz}, and Havara-Lifshitz gravity. It is important to mention here that, in our treatment, we have assumed that all the new effects from spatial covariant gravities are considered as small corrections. In this sense, we observe that the degree of the circular polarization $\Pi$ is small due to the suppressing parameter $|d_1|, \; |d_{-1}| < \mathcal{O}(1)$. In Fig.~\ref{Pi_plot}, we plotted the the analytical expression of the circular polarization $\Pi$ in (\ref{Pi_ana}) (the solid and blue curves) and the corresponding numerical results (the dotted and red curves). From this figure, one can see clearly how well the numerical results are approximated by our analytical ones.

\subsection{Detectability of the parity-violating effects}

As we mentioned in the introduction, the parity-violating effect in PGWs, which is measured by the observable $\Pi$, can produce a lot of observational information in CMB and galaxy surveys. 

In CMB, one important effect is the induction of nonzero the TB and EB spectra in the CMB data. This implies that one can probe the parity violation by measuring the CMB EB and TB angular cross-correlators. However, as analyzed in \cite{parity_power8} (see also \cite{Gerbino:2016mqb}), such proposal is only optimistic when $\Pi > \mathcal{O}(0.5)$, especially considering that the tensor-to-scalar ratio has been constrained to be $r< 0.036$ at 95\% confidence level in Ref. \cite{BICEP:2021xfz}. Note that a more stringent constraint has been reported from a combined analysis of newly released dataset including CMB and GW data \cite{Galloni:2022mok}. Therefore, it seems very difficult to detect these effects in the future CMB experiments. According to the analysis in \cite{Masui:2017fzw},  the main difficulty comes from the two-dimensional projection of CMB, which suppresses the parity-violating  signal due to approximate reflective symmetries, and confuses the tensor modes with scalar ones, leading to additional noise contributions. Possible ways for bypassing such challenge are proposed. The first one is to consider the three-points or even high-order correlators, such as the primordial bispectra and trispectra and their signatures in CMB \cite{Orlando:2022rih, Bartolo:2020gsh}. Another way is to search the tensor fossil effects due to parity violation in the statistics of the large-scale structure in future galaxy surveys \cite{Masui:2017fzw}.  These two topics are obvious out the scope of then current paper and we leave them for our future works. 

Another proposal for detecting primordial parity-violating effects is to consider the imprints of circular polarization on the galaxy intrinsic alignments \cite{Bartolo:2020gsh}. Similar to CMB, the circular polarization $\Pi$ can directly induce a distinctive imprint in the galaxy shape spectrum, i.e., the nonzero EB correlation in the shape spectrum. Considering such signature can not be produced by the scalar modes, any signature of EB correlation  in the future galaxy surveys would be a smoking gun for parity violation in PGWs, as mentioned in \cite{Bartolo:2020gsh}.

\section{conclusion and outlook}
 \renewcommand{\theequation}{6.\arabic{equation}} \setcounter{equation}{0}

The spatial covariant gravities can provide a unified description for a lot of scalar-tensor theories in the  unitary gauge. Such framework breaks the time diffeomorphism of the gravity but respects spatial diffeomorphisms, so that one is able to to include the parity violating odd-order spatial derivative terms but with spatial covariance into the gravitational action.  It is also shown in \cite{Gao:2014soa, Gao:2020yzr} that a lot of parity-violating theories in the unitary gauge can be  mapped to spatial covariant gravities. In this paper, we study the circular polarization of PGWs in the spatial covariant gravities and discuss its possible observational signatures. For this purpose we first solve the evolution of PGWs during slow-roll inflation by applying the uniform asymptotic approximation to the equations of motion for the PGWs. Using this approximation, we construct the approximate analytical solutions to the PGWs during the slow-roll inflation, with which we calculate explicitly both the power spectra for the two polarization modes and the corresponding degree of circular polarization of PGWs in the spatial covariant gravities. It is shown that with the presence of the parity violation, the power spectra of PGWs are slightly modified and the degree of circular polarization becomes nonzero. The magnitude of the degree circular polarization directly depends on the parity-violating terms in the gravitational action (\ref{oldmodel}), which are expected to be quit small due to the suppression of $d_{-1}, d_1 < {\cal O}(1)$. This implies very difficult to detect or effectively constrain the theories by using the TE and EB power spectra of future CMB data. The possible signatures of the circular polarization of PGWs in non-Gaussianities, large-scale structure, and EB correlation in the galaxy-shaped power spectrum are also briefly discussed.

Our work can be improved in several aspects. First, in the current study, we have not yet considered the effects of parity violation arising from the spatial covariant gravities in the non-Gaussianities of PGWs. According to the analysis in \cite{parity_power10}, the parity-violating effects in the tensor-tensor-scalar bispectrum could be large enough and detectable in the future CMB data. Thus, it is interesting to explore further whether the parity-violating terms in spatial covariant gravities could lead to any possible observational signatures in the non-Gaussianity of PGWs. Second, the parity-violating effects in primordial bispectrum and trispectrum of PGWs can also leave imprints in the statistic large scale structure \cite{Masui:2017fzw} as well as the EB correlation in galaxy shape power spectrum \cite{Biagetti:2020lpx}. We would like to come back to these topics in our future works.

\section*{Acknowledgments}

T.Z. and A.W. are supported in part by the National Key Research and Development Program of China Grant No. 2020YFC2201503, and the Zhejiang Provincial Natural Science Foundation of China under Grants No. LR21A050001 and No. LY20A050002, the National Natural Science Foundation of China under Grants No.~12275238, No.~11975203, No.~11675143, and the Fundamental Research Funds for the Provincial Universities of Zhejiang in China under Grant No. RF-A2019015.
W.Z. is supported by NSFC Grants No. 11773028, No. 11633001, No. 11653002, No. 11421303, No. 11903030, the Fundamental Research Funds for the Central Universities, and the Strategic Priority Research Program of the Chinese Academy of Sciences Grant No. XDB23010200. 

\appendix

\section{Calculation of integral of $\sqrt{g}$ in eq. (4.36)}
 \renewcommand{\theequation}{A.\arabic{equation}} \setcounter{equation}{0}

In this appendix, we present the calculation of the following integral,
\bqn
\int_{y}^{y_0} dy' \sqrt{g(y')}.
\eqn
For this purpose, we can write the function $g(y)$ in the following form
\bqn
g(y) = \frac{y_0-y}{y^2}(a_0+a_1 y + a_2 y^2 + a_3 y^3).
\eqn
Here the coefficients $a_0$, $a_1$, $a_2$, and $a_3$ are determined by comparing the above form with (\ref{gofy}), which leads to
\bqn
a_0 &=& d_0 y_0 - d_{-1} \rho_A - d_1 \rho_A y_0^2+ d_2 y_0^3,\\
a_1 &=&d_0 - d_1 \rho_A y_0 + d_2 y_0^2, \\
a_2 &=& -d_1 \rho_A + d_2 y_0,\\
a_3 &=& d_2. 
\eqn
It is evident that the magnitude of the coefficients $a_2$ and $a_3$ are determined by ${\cal O}(d_{1}, d_2)$, which can be treated as small corrections of the new terms beyond GR in spatial covariant gravity. Thus we can expand $\sqrt{g(y)}$ by treating $a_2$ and $a_3$ as small perturbations. Then we have
\bqn
\sqrt{g(y)} &\simeq & \frac{\sqrt{(y_0-y)(a_0+a_1 y)}}{y}  \nb\\
&&+ \frac{1}{2} y (a_2 +a_3 y) \sqrt{\frac{y_0-y}{a_0+a_1 y}}.
\eqn
Thus the integral of $\sqrt{g}$ can be split into two parts,
\bqn
\lim_{y \to 0}\int_{y}^{y_0} dy' \sqrt{g(y')} = I_0 + I_1,
\eqn
where
\bqn
I_0 &=& \sqrt{a_1} y_0 \lim_{x \to 0} \int_{x}^1 \frac{\sqrt{(1-x')(b_0+x')}}{x'} dx' \nb\\
&=& \sqrt{a_1} y_0 \Big[(1-b_0) {\rm arccsc}{(\sqrt{1+b_0})} \nb\\
&&~~~~~~~~+ \sqrt{b_0} {\rm ln} \frac{4b_0}{(1+b_0)x} - \sqrt{b_0}  \Big],  \\
I_1 &=& \frac{a_3 y_0^3}{2\sqrt{a_1}} \lim_{x\to 0}\int_x^1 \sqrt{\frac{1-x'}{b_0+x'}} x' (c_0+x') dx' \nb\\
&=& \frac{a_3 y_0^3}{48\sqrt{a_1}}\Big[\sqrt{b_0} (3-4b_0-15 b_0^2+6c_0+18 b_0 c_0) \nb\\
&&~~~~~~~~~~ + 3(1-2b_0+5b_0^2+2 c_0-6b_0 c_0) \nb\\
&&~~~~~~~~~~~~~~~ \times (1+b_0) {\rm arccsc}{(\sqrt{1+b_0})}\Big],
\eqn
with $x \equiv y/y_0$ and
\bqn
b_0 \equiv \frac{a_0 }{a_1 y_0}, \\
c_0 \equiv \frac{a_2}{a_3 y_0}.
\eqn


\begin{thebibliography}{399}


\bibitem{inflation1}
R. Brout, F. Englert, and E. Gunzig, %
The creation of the universe as a quantum phenomenon, %
Annals Phys. (N.Y.) {\bf 115}, 78 (1978).

\bibitem{inflation2}
K. Sato,
``First Order Phase Transition of a Vacuum and Expansion of the Universe,"
Mon. Not. R. Astron. Soc. {\bf 195}, 467 (1981).

\bibitem{inflation3}
A. H. Guth, %
``The Inflationary Universe: A Possible Solution to the Horizon and Flatness Problems, " %
Phys. Rev. D {\bf 23}, 347 (1981).

\bibitem{inflation4}
A. A. Starobinsky, %
``A new type of isotropic cosmological models without singularity, "%
Phys. Lett. B {\bf 91}, 99 (1980).

\bibitem{inflation5}
A. D. Linde,
 ``A New Inflationary Universe Scenario: A Possible Solution of the Horizon, Flatness, Homogeneity, Isotropy and Primordial Monopole Problems," %
Phys. Lett. B {\bf 108},  389 (1982).

\bibitem{inflation6}
A. Albrecht and P. J. Steinhardt,
``Cosmology for Grand Unified Theories with Radiatively Induced Symmetry Breaking,"
Phys. Rev. Lett. {\bf 48}, 1220 (1982).

\bibitem{Gorski:1996cf}
K. M. Gorski, A. J. Banday, C. L. Bennett, G. Hinshaw, A. Kogut, G. F. Smoot and, E. L. Wright,
``Power spectrum of primordial inhomogeneity determined from the four year COBE DMR sky maps,''
Astrophys. J. {\bf 464}, L11 (1996) [arXiv:astro-ph/9601063].

\bibitem{WMAP}
 WMAP Collaboration, G. Hinshaw et al.,
``Nine-Year Wilkinson Microwave Anisotropy Probe (WMAP) Observations: Cosmological Parameter Results,'' %
Astrophys. J. Suppl. {\bf 208}, 19 (2013) [arXiv:1212.5226 [astro-ph.CO]].

\bibitem{Aghanim:2018eyx}
N. Aghanim  et al. [Planck Collaboration],
``Planck 2018 results. VI. Cosmological parameters,''
arXiv:1807.06209 [astro-ph.CO].

\bibitem{Akrami:2018odb}
Y.~Akrami et al. [Planck Collaboration],
``Planck 2018 results. X. Constraints on inflation,"
arXiv:1807.06211 [astro-ph.CO].


\bibitem{TTEEBB1}
L. M. Krauss, S. Dodelson, and, S. Meyer,
``Primordial Gravitational Waves and Cosmology ,''
Science {\bf 328}, 989 (2010) [arXiv:1004.2504 [astro-ph.CO]].

\bibitem{TTEEBB2}
J. Garcia-Bellido,
``Primordial Gravitational Waves from Inflation and Preheating,''
Prog. Theor. Phys. Suppl. {\bf 190}, 322 (2011) [arXiv:1012.2006 [astro-ph.CO]].

\bibitem{TTEEBB4}
 J. Bock et al.,
``Task Force on Cosmic Microwave Background Research,''
arXiv:astro-ph/ 0604101].

\bibitem{TTEEBB3}
U. Seljak and M. Zaldarriaga,
``Signature of gravity waves in polarization of the microwave background,''
Phys. Rev. Lett. {\bf 78}, 2054 (1997) [arXiv:astro-ph/9609169].

\bibitem{TTEEBB}
M. Kamionkowski, A. Kosowsky and, A. Stebbins,
``A Probe of primordial gravity waves and vorticity ,''
Phys. Rev. Lett. {\bf 78}, 2058 (1997) [arXiv:astro-ph/9609132].


\bibitem{Schmidt:2012nw}
F.~Schmidt and D.~Jeong,
Large-Scale Structure with Gravitational Waves II: Shear,
Phys. Rev. D \textbf{86}, 083513 (2012)
[arXiv:1205.1514 [astro-ph.CO]].


\bibitem{Schmidt:2013gwa}
F.~Schmidt, E.~Pajer and M.~Zaldarriaga,
Large-Scale Structure and Gravitational Waves III: Tidal Effects,
Phys. Rev. D \textbf{89}, 083507 (2014)
[arXiv:1312.5616 [astro-ph.CO]].


\bibitem{Dodelson:2003bv}
S.~Dodelson, E.~Rozo and A.~Stebbins,
Primordial gravity waves and weak lensing,
Phys. Rev. Lett. \textbf{91}, 021301 (2003)
[arXiv:astro-ph/0301177 [astro-ph]].


\bibitem{Dodelson}
S. Dodelson, 
Cross-Correlating Probes of Primordial Gravitational Waves, 
Phys. Rev. D {\bf 82}, 023522 (2010) [arXiv:1001.5012 [astro-ph.CO]].

\bibitem{Schmidt:2012ne}
F.~Schmidt and D.~Jeong,
Cosmic Rulers,
Phys. Rev. D \textbf{86}, 083527 (2012)
[arXiv:1204.3625 [astro-ph.CO]].

\bibitem{Chisari:2014xia}
N.~E.~Chisari, C.~Dvorkin and F.~Schmidt,
Can weak lensing surveys confirm BICEP2?,
Phys. Rev. D \textbf{90}, 043527 (2014)
[arXiv:1406.4871 [astro-ph.CO]].

\bibitem{Biagetti:2020lpx}
M.~Biagetti and G.~Orlando,
Primordial Gravitational Waves from Galaxy Intrinsic Alignments,
JCAP \textbf{07}, 005 (2020)
[arXiv:2001.05930 [astro-ph.CO]].


\bibitem{SimonsObservatory:2018koc}
P.~Ade \textit{et al.} [Simons Observatory],
The Simons Observatory: Science goals and forecasts,
JCAP \textbf{02}, 056 (2019)
[arXiv:1808.07445 [astro-ph.CO]].


\bibitem{Hazumi:2019lys}
M.~Hazumi, P.~A.~R.~Ade, Y.~Akiba, D.~Alonso, K.~Arnold, J.~Aumont, C.~Baccigalupi, D.~Barron, S.~Basak and S.~Beckman, \textit{et al.}
LiteBIRD: A Satellite for the Studies of B-Mode Polarization and Inflation from Cosmic Background Radiation Detection,
J. Low Temp. Phys. \textbf{194}, 443-452 (2019).


\bibitem{CMB-S4:2016ple}
K.~N.~Abazajian \textit{et al.} [CMB-S4],
CMB-S4 Science Book, First Edition, 
arXiv:1610.02743 [astro-ph.CO].

\bibitem{Li:2017drr}
H.~Li, S.~Y.~Li, Y.~Liu, Y.~P.~Li, Y.~Cai, M.~Li, G.~B.~Zhao, C.~Z.~Liu, Z.~W.~Li and H.~Xu, \textit{et al.}, 
Probing Primordial Gravitational Waves: Ali CMB Polarization Telescope,
Natl. Sci. Rev. \textbf{6}, 145-154 (2019)
[arXiv:1710.03047 [astro-ph.CO]].


\bibitem{Amendola:2016saw}
L.~Amendola, S.~Appleby, A.~Avgoustidis, D.~Bacon, T.~Baker, M.~Baldi, N.~Bartolo, A.~Blanchard, C.~Bonvin and S.~Borgani, \textit{et al.}
Cosmology and fundamental physics with the Euclid satellite,
Living Rev. Rel. \textbf{21}, 2 (2018)
[arXiv:1606.00180 [astro-ph.CO]].


\bibitem{LSST:2008ijt}
\v{Z}.~Ivezi\'c \textit{et al.} [LSST],
LSST: from Science Drivers to Reference Design and Anticipated Data Products,
Astrophys. J. \textbf{873}, 111 (2019)
[arXiv:0805.2366 [astro-ph]].



\bibitem{parity_CMB1}
 N. Seto and A. Taruya,
 ``Measuring a Parity Violation Signature in the Early Universe via Ground-based Laser Interferometers,''
Phys. Rev. Lett. {\bf 99}, 121101 (2007) [arXiv:0707.0535 [astro-ph]].

\bibitem{parity_CMB}
A. Lue, L. Wang and, M. Kamionkowski,
``Cosmological signature of new parity violating interactions,''
Phys. Rev. Lett. {\bf 83}, 1506 (1999) [arXiv:astro-ph/9812088].

\bibitem{parity_CMB2}
S. Saito, K. Ichiki, and, A. Taruya,
``Probing polarization states of primordial gravitational waves with CMB anisotropies,''
J. Cosmol. Astropart. Phys. {\bf 09}, 002 (2007) [arXiv:0705.3701 [astro-ph]].

\bibitem{parity_CMB3}
 C. Bischoff et al. Quiet Collaboration,
``First Season QUIET Observations: Measurements of CMB Polarization Power Spectra at 43 GHz in the Multipole Range $25\leq \ell \leq 475$ ,''
Astrophys. J. {\bf 741}, 111 (2011) [arXiv:1012.3191 [astro-ph.CO]].

\bibitem{parity_CMB4}
V. Gluscevic and M. Kamionkowski,
``Testing Parity-Violating Mechanisms with Cosmic Microwave Background Experiments,''
Phys. Rev. D {\bf 81}, 123529 (2010) [arXiv:1002.1308 [astro-ph.CO]].

\bibitem{Seto:2007tn}
N.~Seto and A.~Taruya,
Measuring a Parity Violation Signature in the Early Universe via Ground-based Laser Interferometers,
Phys. Rev. Lett. \textbf{99}, 121101 (2007)
[arXiv:0707.0535 [astro-ph]].

\bibitem{Seto:2008sr}
N.~Seto and A.~Taruya,
Polarization analysis of gravitational-wave backgrounds from the correlation signals of ground-based interferometers: Measuring a circular-polarization mode,
Phys. Rev. D \textbf{77}, 103001 (2008)
[arXiv:0801.4185 [astro-ph]].

\bibitem{Seto:2020zxw}
N.~Seto,
Measuring Parity Asymmetry of Gravitational Wave Backgrounds with a Heliocentric Detector Network in the mHz Band,
Phys. Rev. Lett. \textbf{125}, 251101 (2020)
[arXiv:2009.02928 [gr-qc]].

\bibitem{Orlando:2020oko}
G.~Orlando, M.~Pieroni and A.~Ricciardone,
Measuring Parity Violation in the Stochastic Gravitational Wave Background with the LISA-Taiji network,
JCAP \textbf{03}, 069 (2021)
[arXiv:2011.07059 [astro-ph.CO]].


\bibitem{Seto:2006dz}
N.~Seto,
Quest for circular polarization of gravitational wave background and orbits of laser interferometers in space,
Phys. Rev. D \textbf{75}, 061302 (2007)
[arXiv:astro-ph/0609633 [astro-ph]].

\bibitem{Masui:2017fzw}
K.~W.~Masui, U.~L.~Pen and N.~Turok,
Two- and Three-Dimensional Probes of Parity in Primordial Gravity Waves,
Phys. Rev. Lett. \textbf{118}, 221301 (2017)
[arXiv:1702.06552 [astro-ph.CO]].


\bibitem{Lue:1998mq}
A.~Lue, L.~M.~Wang and M.~Kamionkowski,
Cosmological signature of new parity violating interactions,
Phys. Rev. Lett. \textbf{83}, 1506 (1999)
[arXiv:astro-ph/9812088 [astro-ph]].

\bibitem{Alexander:2009tp}
S.~Alexander and N.~Yunes,
Chern-Simons Modified General Relativity,
Phys. Rept. \textbf{480}, 1-55 (2009)
[arXiv:0907.2562 [hep-th]].


\bibitem{Crisostomi:2017ugk}
M.~Crisostomi, K.~Noui, C.~Charmousis and D.~Langlois,
Beyond Lovelock gravity: Higher derivative metric theories,
Phys. Rev. D \textbf{97}, 044034 (2018)
[arXiv:1710.04531 [hep-th]].


\bibitem{parity_power8}
A. Wang, Q. Wu, W. Zhao and, T. Zhu,
``Polarizing primordial gravitational waves by parity violation,''
Phys. Rev. D {\bf 87}, 103512 (2013) [arXiv:1208.5490 [astro-ph.CO]].

\bibitem{Takahashi:2009wc}
T. Takahashi and J. Soda,
``Chiral Primordial Gravitational Waves from a Lifshitz Point,''
Phys. Rev. Lett. {\bf 102}, 231301 (2009) [arXiv:0904.0554 [hep-th]].

\bibitem{Zhu:2013fja}
T. Zhu, W. Zhao, Y. Huang, A. Wang and, Q. Wu,
``Effects of parity violation on non-gaussianity of primordial gravitational waves in Ho\v{r}ava-Lifshitz gravity,''
Phys. Rev. D {\bf 88}, 063508 (2013) [arXiv:1305.0600 [hep-th]].

\bibitem{Wang:2017brl}
A.~Wang, 
Ho\v{r}ava gravity at a Lifshitz point: A progress report,
Int. J. Mod. Phys. D \textbf{26}, 1730014 (2017)
[arXiv:1701.06087 [gr-qc]].


\bibitem{Gao:2014soa}
X.~Gao,
Unifying framework for scalar-tensor theories of gravity,
Phys. Rev. D \textbf{90}, 081501 (2014)
[arXiv:1406.0822 [gr-qc]].

\bibitem{Gao:2014fra}
X.~Gao,
Hamiltonian analysis of spatially covariant gravity,
Phys. Rev. D \textbf{90}, 104033 (2014)
[arXiv:1409.6708 [gr-qc]].

\bibitem{Gao:2018znj}
X.~Gao and Z.~B.~Yao,
Spatially covariant gravity with velocity of the lapse function: the Hamiltonian analysis,
\JCAP \textbf{05}, 024 (2019)
[arXiv:1806.02811 [gr-qc]].

\bibitem{Gao:2019lpz}
X.~Gao, C.~Kang and Z.~B.~Yao,
Spatially Covariant Gravity: Perturbative Analysis and Field Transformations,
Phys. Rev. D \textbf{99}, 104015 (2019)
[arXiv:1902.07702 [gr-qc]].

\bibitem{Gao:2020yzr}
X.~Gao and Y.~M.~Hu,
Higher derivative scalar-tensor theory and spatially covariant gravity: the correspondence,
Phys. Rev. D \textbf{102}, 084006 (2020)
[arXiv:2004.07752 [gr-qc]].


\bibitem{Li:2020xjt}
M.~Li, H.~Rao and D.~Zhao,
``A simple parity violating gravity model without ghost instability,''
JCAP \textbf{11}, 023 (2020)
[arXiv:2007.08038 [gr-qc]].


\bibitem{Li:2021wij}
M.~Li, H.~Rao and Y.~Tong,
Revisit on the healthy parity violating gravity model: local Lorentz covariance,
[arXiv:2104.05917 [gr-qc]].

\bibitem{Conroy:2019ibo}
A.~Conroy and T.~Koivisto,
Parity-Violating Gravity and GW170817 in Non-Riemannian Cosmology,
JCAP \textbf{12}, 016 (2019)
[arXiv:1908.04313 [gr-qc]].

\bibitem{Li:2022vtn}
M.~Li, Y.~Tong and D.~Zhao,
Possible consistent model of parity violations in the symmetric teleparallel gravity,
Phys. Rev. D \textbf{105}, 104002 (2022)
[arXiv:2203.06912 [gr-qc]].


\bibitem{Kostelecky:2016kfm}
V.~A.~Kosteleck\'y and M.~Mewes,
Testing local Lorentz invariance with gravitational waves,
Phys. Lett. B \textbf{757}, 510-514 (2016)
[arXiv:1602.04782 [gr-qc]].

\bibitem{Bailey:2006fd}
Q.~G.~Bailey and V.~A.~Kostelecky,
Signals for Lorentz violation in post-Newtonian gravity,
Phys. Rev. D \textbf{74}, 045001 (2006)
[arXiv:gr-qc/0603030 [gr-qc]].


\bibitem{Mewes:2019dhj}
M.~Mewes,
Signals for Lorentz violation in gravitational waves,
Phys. Rev. D \textbf{99}, 104062 (2019)
[arXiv:1905.00409 [gr-qc]].

\bibitem{Shao:2020shv}
L.~Shao,
Combined search for anisotropic birefringence in the gravitational-wave transient catalog GWTC-1,
Phys. Rev. D \textbf{101}, 104019 (2020)
[arXiv:2002.01185 [hep-ph]].

\bibitem{Wang:2021ctl}
Z.~Wang, L.~Shao and C.~Liu,
New Limits on the Lorentz/CPT Symmetry Through 50 Gravitational-wave Events,
Astrophys. J. \textbf{921}, 158 (2021)
[arXiv:2108.02974 [gr-qc]].

\bibitem{Contaldi:2008yz}
C.~R.~Contaldi, J.~Magueijo and L.~Smolin,
Anomalous CMB polarization and gravitational chirality,
Phys. Rev. Lett. \textbf{101}, 141101 (2008)
[arXiv:0806.3082 [astro-ph]].

\bibitem{Qiao:2021fwi}
J.~Qiao, T.~Zhu, G.~Li and W.~Zhao,
Post-Newtonian parameters of ghost-free parity-violating gravities,
JCAP \textbf{04}, 054 (2022)
[arXiv:2110.09033 [gr-qc]].

\bibitem{Qiao:2019wsh}
J.~Qiao, T.~Zhu, W.~Zhao and A.~Wang,
Waveform of gravitational waves in the ghost-free parity-violating gravities,
Phys. Rev. D \textbf{100}, 124058 (2019)
[arXiv:1909.03815 [gr-qc]].


\bibitem{Zhang:2022xmm}
F.~Zhang, J.~X.~Feng and X.~Gao,
Circularly polarized scalar induced gravitational waves from the Chern-Simons modified gravity,
[arXiv:2205.12045 [gr-qc]].

\bibitem{Orlando:2022rih}
G.~Orlando,
Probing parity-odd bispectra with anisotropies of GW $V$ modes,
arXiv:2206.14173 [astro-ph.CO].

\bibitem{Chen:2022soq}
J.~Chen, S.~Ghosh and W.~Zhao,
Scalar Quadratic Maximum-likelihood Estimators for the CMB Cross-power Spectrum,
Astrophys. J. Supp. \textbf{260}, 44 (2022)
[arXiv:2202.10733 [astro-ph.IM]].

\bibitem{Cai:2021uup}
R.~G.~Cai, C.~Fu and W.~W.~Yu,
Parity violation in stochastic gravitational wave background from inflation in Nieh-Yan modified teleparallel gravity,
Phys. Rev. D \textbf{105}, 103520 (2022)
[arXiv:2112.04794 [astro-ph.CO]].

\bibitem{Fronimos:2021czc}
F.~P.~Fronimos and S.~A.~Venikoudis,
Inflation with exotic kinetic terms in Einstein\textendash{}Chern\textendash{}Simons gravity,
Int. J. Mod. Phys. A \textbf{36}, 2150229 (2021)
[arXiv:2110.12457 [gr-qc]].

\bibitem{Bartolo:2020gsh}
N.~Bartolo, L.~Caloni, G.~Orlando and A.~Ricciardone,
Tensor non-Gaussianity in chiral scalar-tensor theories of gravity,
JCAP \textbf{03}, 073 (2021)
[arXiv:2008.01715 [astro-ph.CO]].

\bibitem{Fu:2020tlw}
C.~Fu, J.~Liu, T.~Zhu, H.~Yu and P.~Wu,
``Resonance instability of primordial gravitational waves during inflation in Chern\textendash{}Simons gravity,''
Eur. Phys. J. C \textbf{81}, 204 (2021)
[arXiv:2006.03771 [gr-qc]].

\bibitem{Mylova:2019jrj}
M.~Mylova,
``Chiral primordial gravitational waves in extended theories of Scalar-Tensor gravity,''
[arXiv:1912.00800 [gr-qc]].

\bibitem{parity_power10}
N. Bartolo and G. Orlando,
Parity breaking signatures from a Chern-Simons coupling during inflation: the case of non-Gaussian gravitational waves,
J. Cosmol. Astropart. Phys. {\bf 07}, 034 (2017) [arXiv:1706.04627 [astro-ph.CO]].

\bibitem{Hu:2020rub}
Q.~Hu, M.~Li, R.~Niu and W.~Zhao,
Joint Observations of Space-based Gravitational-wave Detectors: Source Localization and Implication for Parity-violating gravity,
Phys. Rev. D \textbf{103}, 064057 (2021)
[arXiv:2006.05670 [gr-qc]].

\bibitem{Odintsov:2021kup}
S.~D.~Odintsov, V.~K.~Oikonomou and F.~P.~Fronimos,
Quantitative predictions for f(R) gravity primordial gravitational waves,
Phys. Dark Univ. \textbf{35}, 100950 (2022)
[arXiv:2108.11231 [gr-qc]].

\bibitem{Odintsov2}
S. D. Odintsov and V. K. Oikonomou, 
Chirality of Gravitational Waves in Chern-Simons f(R) Gravity Cosmology, 
Phys. Rev. D {\bf 105}, 104054 (2022).  

\bibitem{Peng:2022ttg}
Z.~Z.~Peng, Z.~M.~Zeng, C.~Fu and Z.~K.~Guo,
Generation of gravitational waves in dynamical Chern-Simons gravity,
[arXiv:2209.10374 [gr-qc]].

\bibitem{Kamada:2021kxi}
K.~Kamada, J.~Kume and Y.~Yamada,
Chiral gravitational effect in time-dependent backgrounds,
JHEP \textbf{05}, 292 (2021)
[arXiv:2104.00583 [hep-ph]].


\bibitem{Lyth:2005jf}
D.~H.~Lyth, C.~Quimbay and Y.~Rodriguez,
Leptogenesis and tensor polarisation from a gravitational Chern-Simons term,
JHEP \textbf{03}, 016 (2005)
[arXiv:hep-th/0501153 [hep-th]].

\bibitem{Nilsson:2022mzq}
N.~A.~Nilsson,
``Explicit spacetime-symmetry breaking and the dynamics of primordial fields,''
Phys. Rev. D \textbf{106}, 104036 (2022)
[arXiv:2205.00496 [gr-qc]].

\bibitem{soda_JHEP}
J.~Soda, H.~Kodama and M.~Nozawa,
``Parity Violation in Graviton Non-gaussianity,''
JHEP \textbf{08} (2011), 067
[arXiv:1106.3228 [hep-th]].

\bibitem{soda_JCAP}
M.~Satoh and J.~Soda,
``Higher Curvature Corrections to Primordial Fluctuations in Slow-roll Inflation,''
JCAP \textbf{09} (2008), 019
[arXiv:0806.4594 [astro-ph]].

\bibitem{soda_PRD}
M.~Satoh, S.~Kanno and J.~Soda,
``Circular Polarization of Primordial Gravitational Waves in String-inspired Inflationary Cosmology,''
Phys. Rev. D \textbf{77} (2008), 023526
[arXiv:0706.3585 [astro-ph]].

\bibitem{Qiao:2022mln}
J.~Qiao, Z.~Li, T.~Zhu, R.~Ji, G.~Li and W.~Zhao,
``Testing parity symmetry of gravity with gravitational waves,''
Front. Astron. Space Sci. {\bf 9}, 1109086 (2023).
[arXiv:2211.16825 [gr-qc]].

\bibitem{Qiao:2019hkz}
J.~Qiao, T.~Zhu, W.~Zhao and A.~Wang,
``Polarized primordial gravitational waves in the ghost-free parity-violating gravity,''
Phys. Rev. D \textbf{101},  043528 (2020)
[arXiv:1911.01580 [astro-ph.CO]].

\bibitem{Gao:2019liu}
X.~Gao and X.~Y.~Hong,
Propagation of gravitational waves in a cosmological background,
Phys. Rev. D \textbf{101}, 064057 (2020)
[arXiv:1906.07131 [gr-qc]].


\bibitem{Zhao:2019xmm}
W.~Zhao, T.~Zhu, J.~Qiao and A.~Wang,
Waveform of gravitational waves in the general parity-violating gravities,
Phys. Rev. D \textbf{101}, 024002 (2020)
[arXiv:1909.10887 [gr-qc]].

\bibitem{Wang:2020cub}
Y.~F.~Wang, R.~Niu, T.~Zhu and W.~Zhao,
Gravitational Wave Implications for the Parity Symmetry of Gravity in the High Energy Region,
Astrophys. J. \textbf{908}, 58 (2021)
[arXiv:2002.05668 [gr-qc]].



\bibitem{Okounkova:2021xjv}
M.~Okounkova, W.~M.~Farr, M.~Isi and L.~C.~Stein,
Constraining gravitational wave amplitude birefringence and Chern-Simons gravity with GWTC-2,
Phys. Rev. D {\bf 106}, 044067 (2022) 
[arXiv: 2101.11153 [gr-qc]].



\bibitem{Nishizawa:2018srh}
A.~Nishizawa and T.~Kobayashi,
Parity-violating gravity and GW170817,
Phys. Rev. D \textbf{98}, 124018 (2018)
[arXiv:1809.00815 [gr-qc]].


\bibitem{Zhao:2019szi}
W.~Zhao, T.~Liu, L.~Wen, T.~Zhu, A.~Wang, Q.~Hu and C.~Zhou,
Model-independent test of the parity symmetry of gravity with gravitational waves,
Eur. Phys. J. C \textbf{80}, 630 (2020)
[arXiv:1909.13007 [gr-qc]].


%
\bibitem{Wang:2017igw}
S.~Wang,
Exploring the CPT violation and birefringence of gravitational waves with ground- and space-based gravitational-wave interferometers,
Eur. Phys. J. C \textbf{80}, 342 (2020)
[arXiv:1712.06072 [gr-qc]].

\bibitem{Wang:2021gqm}
Y.~F.~Wang, S.~M.~Brown, L.~Shao and W.~Zhao,
Tests of Gravitational-Wave Birefringence with the Open Gravitational-Wave Catalog,
Phys. Rev. D {\bf 106}, 084005 (2022) 
[arXiv:2109.09718 [astro-ph.HE]].

\bibitem{Zhao:2022pun}
Z.~C.~Zhao, Z.~Cao and S.~Wang,
Search for the Birefringence of Gravitational Waves with the Third Observing Run of Advanced LIGO-Virgo,
Astrophys. J. \textbf{930}, 139 (2022)
[arXiv:2201.02813 [gr-qc]].

\bibitem{Wang:2020pgu}
S.~Wang and Z.~C.~Zhao,
Tests of CPT invariance in gravitational waves with LIGO-Virgo catalog GWTC-1,
Eur. Phys. J. C \textbf{80}, 1032 (2020)
[arXiv:2002.00396 [gr-qc]].

\bibitem{Niu:2022yhr}
R.~Niu, T.~Zhu and W.~Zhao,
Constraining Anisotropy Birefringence Dispersion in Gravitational Wave Propagation with GWTC-3,
[arXiv:2202.05092 [gr-qc]].

\bibitem{Gong:2021jgg}
C.~Gong, T.~Zhu, R.~Niu, Q.~Wu, J.~L.~Cui, X.~Zhang, W.~Zhao and A.~Wang,
Gravitational wave constraints on Lorentz and parity violations in gravity: High-order spatial derivative cases,
Phys. Rev. D \textbf{105}, 044034 (2022)
[arXiv:2112.06446 [gr-qc]].

\bibitem{Wu:2021ndf}
Q.~Wu, T.~Zhu, R.~Niu, W.~Zhao and A.~Wang,
Constraints on the Nieh-Yan modified teleparallel gravity with gravitational waves,
Phys. Rev. D \textbf{105}, 024035 (2022)
[arXiv:2110.13870 [gr-qc]].


\bibitem{olver1954}
F. W. J. Olver, Asymptotics and Special functions (AKP Classics, Wellesley, MA, 1997).

\bibitem{zhu_inflationary_2014}%
T. Zhu, A. Wang, G. Cleaver, K. Kirsten and, Q. Sheng,
``Inflationary cosmology with nonlinear dispersion relations ,''
Phys. Rev. D {\bf 89}, 043507 (2014) [arXiv:1308.5708[astro-ph.CO]].


\bibitem{qiao_inflationary_2018}
J. Qiao, G. Ding, Q. Wu, T. Zhu and, A. Wang, %
``Inflationary perturbation spectrum in extended effective field theory of inflation,''
J. Cosmol. Astropart. Phys. {\bf 09}, 064 (2019) [arXiv:1811.03216 [hep-th]].



\bibitem{habib_inflationary_2002}
S. Habib, K. Heitmann, G. Jungman and, C. Molina-Paris,
``The Inflationary perturbation spectrum,''
Phys. Rev. Lett. {\bf 89}, 281301 (2002) [arXiv:astro-ph/0208443].


\bibitem{zhu_power_2014}%
T. Zhu, A. Wang, G. Cleaver, K. Kirsten, and Q. Sheng,
``Power spectra and spectral indices of $k$-inflation: high-order corrections,''
Phys. Rev. D {\bf 90}, 103517 (2014) [arXiv:1407.8011 [astro-ph.CO]].


\bibitem{zhu_constructing_2014}%
T. Zhu, A. Wang, G. Cleaver, K. Kirsten, and Q. Sheng,
``Constructing analytical solutions of linear perturbations of inflation with modified dispersion relations,''
Int. J. Mod. Phys. A {\bf 29}, 1450142 (2014) [arXiv:1308.1104 [astro-ph.CO]].

\bibitem{Zhu:2018smk}
T.~Zhu, Q.~Wu and, A.~Wang,
``An analytical approach to the field amplification and particle production by parametric resonance during inflation and reheating,''
Phys.\ Dark Univ.\  {\bf 26}, 100373 (2019) [arXiv:1811.12612[hep-ph]].

\bibitem{Zhu:2019bwj}
T. Zhu and A. Wang,
``Langer Modification, Quantization condition and Barrier Penetration in Quantum Mechanics,''
Universe {\bf 6}, 90 (2020) [arXiv:1902.09675[quant-ph]].

\bibitem{Gerbino:2016mqb}
M.~Gerbino, A.~Gruppuso, P.~Natoli, M.~Shiraishi and A.~Melchiorri,
Testing chirality of primordial gravitational waves with Planck and future CMB data: no hope from angular power spectra,
JCAP \textbf{07}, 044 (2016)
[arXiv:1605.09357 [astro-ph.CO]].

\bibitem{BICEP:2021xfz}
P.~A.~R.~Ade \textit{et al.} [BICEP and Keck],
Improved Constraints on Primordial Gravitational Waves using Planck, WMAP, and BICEP/Keck Observations through the 2018 Observing Season,
Phys. Rev. Lett. \textbf{127}, 151301 (2021)
[arXiv:2110.00483 [astro-ph.CO]].

\bibitem{Galloni:2022mok}
G.~Galloni, N.~Bartolo, S.~Matarrese, M.~Migliaccio, A.~Ricciardone and N.~Vittorio,
Updated constraints on amplitude and tilt of the tensor primordial spectrum,
[arXiv:2208.00188 [astro-ph.CO]].


\end{thebibliography}
\end{document}